\documentstyle[psfig]{mn}

\input{epsf}

\def\lsim{\mathrel{\hbox{\rlap{\hbox{\lower4pt\hbox{$\sim$}}}\hbox{$<$}}}}
\def\gsim{\mathrel{\hbox{\rlap{\hbox{\lower4pt\hbox{$\sim$}}}\hbox{$>$}}}}

\def\and {\rm {et al.} \rm} 
\def\etal {\rm {et al.} \rm}

\begin{document}

\title[The power spectrum of galaxy clustering in the APM survey]
{The power spectrum of galaxy clustering in the APM Survey}

\author[N. D. Padilla \& C. M. Baugh]
{N. D. Padilla and C. M. Baugh
\\
Department of Physics, University of Durham, 
Science Laboratories, South Road, Durham DH1 3LE \\
}

\maketitle 
 
\begin{abstract}
We measure the power spectrum of galaxy clustering in real 
space from the APM Galaxy Survey. We present an improved 
technique for the numerical inversion of Limber's equation that 
relates the angular clustering of galaxies to an integral over 
the power spectrum in three dimensions. Our approach is 
underpinned by a large ensemble of mock galaxy catalogues 
constructed from the Hubble Volume N-body simulations.
The mock catalogues are used to test for systematic effects 
in the inversion algorithm and to estimate the errors on our 
measurement. 
We find that we can recover the power spectrum to an accuracy 
of better than $15\%$ over three decades in wavenumber. 
A key advantage of the use of mock catalogues to infer errors is  
that we can apply our technique on scales for which the density 
fluctuations are not Gaussian, thus probing the regime that 
offers the best constraints on models of galaxy formation.
On large scales, our measurement of the power spectrum is 
consistent with the shape of the mass power spectrum in the 
popular ``concordance'' cold dark matter model. The galaxy 
power spectrum on small scales is strongly affected by nonlinear 
evolution of density fluctuations, and, to a lesser degree, by 
galaxy bias. 
The {\it rms} variance in the galaxy distribution, when smoothed 
in spheres of radius $8h^{-1}$Mpc, is $\sigma^{g}_8=0.96^{+0.17}_{-0.20}$ 
and the shape of the power spectrum on large scales 
is described by a simple fitting 
formula with parameter $\Gamma= 0.19^{+0.13}_{-0.04}$ (these errors are 
the $1 \sigma$ ranges for a two parameter fit). 
We use our measurement of the power spectrum to estimate the galaxy two 
point correlation function; the results are well described by a power 
law with correlation length $r_0 = 5.9 \pm 0.7 h^{-1}$Mpc and slope 
$\gamma = 1.61 \pm 0.06$ 
for pair separations in the range $0.1 < r/(h^{-1}{\rm Mpc}) < 20$. 
\end{abstract}

\begin{keywords}
methods: statistcial - methods: numerical -
large-scale structure of Universe - galaxies: formation
\end{keywords}

\section{Introduction}

The power spectrum of galaxy clustering is the equivalent of the 
Rosetta stone for structure formation in the Universe.
On large scales, the power spectrum is a fossil record of the 
density fluctuations present at the epoch of recombination. 
Moving to intermediate and small scales, the spectrum encodes a wealth of 
information about different physical processes and effects: 
the nonlinear evolution of the density field, the modulation of the 
clustering signal of galaxies relative to that of the underlying dark 
matter, and, if the spectrum is estimated using the positions of 
galaxies inferred from their redshifts, the magnitude of peculiar 
motions and bulk flows.

In the past decade, significant progress has been made in measuring 
the power spectrum of galaxy clustering. Two main approaches have been 
taken: direct estimation in three dimensions using galaxy redshift 
surveys (e.g. Tadros \& Efstathiou 1996; Hoyle \etal 1999; 
Sutherland \etal 1999; Percival \etal 2001) 
and the deprojection of angular clustering (e.g. Peacock 1991; Baugh \& 
Efstathiou 1993, 1994a). Until the advent of the PSCz (Saunders \etal 2000) 
and 2dF Galaxy Redshift Survey (2dFGRS, Colless \etal 2001), 
angular catalogues (e.g. the APM Survey 
described by Maddox, Efstathiou \& Sutherland, 1996) 
generally covered a much larger volume 
than that mapped out by the contemporaneous redshift surveys. 
On completion of the photometric part of the Sloan Digital Sky Survey (SDSS), 
which is deeper than the spectroscopic catalogue, this will once 
again be the case. With photometric redshifts it will be possible to 
isolate SDSS galaxies in slices in redshift and to apply a deprojection 
algorithm to the angular clustering measured in each slice in order 
to quantify the evolution of the power spectrum (Dodelson \etal 2002). 
Other new angular catalogues are now becoming available.
The Two Micron All-Sky Survey (2MASS) has mapped the whole sky in 
the near-infrared, covering a solid angle four times larger than the SDSS, 
albeit to a shallower depth (Jarrett \etal 2000). 
(A preliminary estimate of the power spectrum of 2MASS galaxies was  
made by Allgood, Blumenthal \& Primack 2001. After our paper was refereed, 
Maller \etal (2003) produced a detailed analysis of the power spectrum of 
2MASS galaxies in three dimensions, using the completed survey.)
Deeper photometric catalogues covering relatively large solid angles are 
also being constructed, which will allow measurements of the power 
spectrum of galaxy clustering to be undertaken at high 
redshift (e.g. McCracken \etal 2001).

The motivation to develop reliable algorithms to estimate 
the power spectrum in three dimensions from projected clustering data 
is therefore clear.
Baugh \& Efstathiou (1993, hereafter BE93) introduced a technique 
to invert a variant of Limber's equation (Limber 1954), which relates the angular 
correlation function, $w(\theta)$, to an integral over the power 
spectrum, $P(k)$. The inversion scheme is based upon a Bayesian procedure 
proposed by Lucy (1974). 
An advantage of this approach is that it does not depend upon any prior 
prejudice about the form of the power spectrum. 
On the other hand, a disadvantage of the method is that it does not 
yield the error on the recovered power spectrum directly. 
Baugh \& Efstathiou estimated the error on their measurement 
by dividing the APM survey into four zones, applying the inversion to 
the clustering measured in each zone  and using the scatter between 
these estimates to give the errors.
The inversion method was also applied to the angular power spectrum to infer 
the power spectrum in three dimensions (Baugh \& Efstathiou 1994) and to 
Limber's original equation to obtain the spatial correlation function, 
$\xi(r)$ (Baugh 1996). Gazta\~{n}aga \& Baugh (1998, hereafter GB98) 
presented tests of the 
algorithm using mock catalogues constructed from N-body simulations. 

Several recent papers have proposed new algorithms to invert measurements 
of projected clustering using matrix techniques which also produce an 
estimate of the covariance matrix for the power spectrum.
A direct inversion of Limber's equation is unstable so various 
approximations have to be made to overcome this.
Dodelson \& Gazta\~{n}aga (2000) adopted a Bayesian prior relating to the 
smoothness of the power spectrum to stabilise the inversion. The magnitude of 
the estimated errors is somewhat sensitive to the choice of smoothing 
parameter. These authors also assumed a diagonal covariance matrix and 
a Gaussian distribution of errors for the angular correlation function.
Eisenstein \& Zaldarriaga (2001) proposed a singular-value decomposition of 
Limber's equation written in matrix form, without smoothing. These authors 
used a fit to the power spectrum in two dimensions measured from the 
APM by Baugh \& Efstathiou (1994), and assumed a Gaussian distribution of 
errors to compute the covariance matrix on this quantity. 
Efstathiou \& Moody (2001) devised a maximum likelihood estimator for 
the power spectrum in three dimensions based upon the measured projected 
counts of galaxies in cells. Again, the fluctuations in cell counts were 
assumed to be Gaussian. In these papers, the assumption of Gaussianity 
limits the range of scales over which the power spectrum can be recovered.

The goal of this paper is to present a reliable measurement of the power 
spectrum in three dimensions, with a robust estimate of the errors 
on our measurement, using the APM Survey as an example. 
This paper contains a number of advances over previous work: 

\begin{itemize}
\item[(i)] We present a revised version of the Lucy algorithm used by 
Baugh \& Efstathiou (1993). Lucy's algorithm has the appeal of simplicity 
and the iterations converge to the maximum likelihood solution to the 
inversion of the integral equation (see Lucy 1974, 1994).

\item[(ii)] The assumption of Gaussian distributed 
density fluctuations is not required for the 
inversion to work. This means that we can estimate the power spectrum on 
small scales where the assumption of Gaussian fluctuations is a poor one 
(e.g. Baugh, Gazta\~{n}aga \& Efstathiou 1995). This allows us to measure 
the power spectrum in a regime that provides a powerful 
constraint on models of galaxy formation.

\item[(iii)] We employ mock catalogues extracted from large N-body 
simulations to make a robust estimate of the errors on the recovered 
power spectrum and of the correlation between the measurements at 
different wavenumbers. We note that Dodelson \etal (2002) have also used 
artificial catalogues 
to estimate the errors on the power spectrum recovered from an 
early version of the SDSS photometric catalogue. The only assumption 
we make in our error estimation is that the mock catalogues provide 
a realistic representation of the sample variance applicable to the case
of the APM survey in the real Universe.

\item[(iv)] We use a new model for the galaxy redshift distribution 
from the 2dFGRS (Colless \etal 2001) based upon the survey selection 
function derived by Norberg \etal (2002a).
\end{itemize}

In Section 2 we give an overview of the inversion algorithm, presenting 
tests of the method using mock catalogues and including a comparison of our 
error estimates with those of other authors. The algorithm is applied 
to the APM Survey in Section 3 and the constraints on the parameters of the 
cold dark matter model from the measurement of the galaxy 
power spectrum are presented in Section 4. We infer the galaxy correlation 
function, $\xi(r)$, from the power spectrum in Section 5. Finally, the 
conclusions are given in Section 6.

\section{Inversion of Limber's Equation}
\label{sec:inversion}

In this section, we describe the numerical inversion of 
Limber's equation. This section is relatively long and detailed, 
so we have split it up into a number of subsections as follows. 
The relativistic form of the integral equation relating the 
angular correlation function, $w(\theta)$, to the power spectrum 
in three dimensions, $P(k)$, is given in Section \ref{ssec:limber}. 
The basic principles of the numerical inversion of this integral equation 
are set out in Section \ref{ssec:lucy}. 
The modifications to the algorithm of Baugh \& Efstathiou (1993) that 
we employ in this paper are explained in Section \ref{ssec:stopping}.
The evolution of clustering over the redshift interval 
probed by the APM Survey is discussed in Section \ref{ssec:ev}. 
The construction of the mock catalogues used to test the method 
and to estimate the errors on the recovered spectrum is reviewed in Section 
\ref{ssec:mocks}.
The inversion of Limber's equation requires the redshift distribution of 
galaxies to be specified. The form used for the redshift distribution of 
APM Survey galaxies is justified in Section \ref{ssec:nofz}.
The role of the mock catalogues in the estimation of the errors 
on the power spectrum recovered from the inversion is explained 
in Section \ref{ssec:error_estimation}.
The numerical inversion is tested using the Hubble Volume mock 
catalogues in Section \ref{ssec:app_mocks}.
Finally, we close the section by comparing our estimate of the error on 
$P(k)$ with others in the literature in Section \ref{ssec:error_comp}.

\subsection{Limber's equation}
\label{ssec:limber}

The angular correlation function, $w(\theta)$, is related
to the real space power spectrum, $P(k)$, through a modified version 
of Limber's  equation (Limber 1954),
\begin{equation}
w(\omega)=\int_0^{\infty}P(k)k g(k \omega){\rm d}k,
\label{eq:limber}
\end{equation}
where $\omega=2 \sin(\theta/2)$, and the kernel function,
$g(k \omega)$, is defined by (BE93,
note that BE93 has a typo in this equation): 
\begin{equation}
g(k \omega) = \frac1{2\pi C} 
\int_0^{\infty}
\frac{F(x)}{(1+z)^{\alpha}} \left( \frac{{\rm d}N}{{\rm d}z} \right)^2
z\frac{{\rm d}z}{{\rm d}x} J_0(k\omega x) {\rm d}z,
\label{eq:kernel}
\end{equation}
where $C = ({\cal N}\Omega_s)^2$, $\Omega_s$ is the solid angle 
of the survey, ${\cal N}$ is the surface density of galaxies on the 
sky, $x$ is the comoving distance, 
and ${\rm d}N/{\rm d}z$ is the redshift 
distribution of galaxies. 
The term $F(x)$ depends on the cosmological model (see e.g.
Peebles 1980; Peebles 1993): 
\begin{equation}
1/{F(x)} = \sqrt{\Omega_{m}(1+z)^{3} + 1 -\Omega_{m}},
\end{equation}
where $\Omega_{m}$ is the present day matter density parameter.

The evolution of the power spectrum with redshift is parameterised 
as 
\begin{equation}
P(k,z)=P(k,z=0)/(1+z)^{\alpha}, 
\label{eq:alpha}
\end{equation}
where $k$ is the comoving wavenumber and $\alpha$ is a parameter.
This is undoubtedly a simplification but one that is justified 
by the relatively shallow depth of the APM survey, $z_m \simeq 0.13$ 
(see Section \ref{ssec:nofz}).
We examine the accuracy of this model in Section \ref{ssec:ev},  
in which we also motivate our choice of value for the parameter 
$\alpha$.

\subsection{Inversion of Limber's equation using Lucy's method}
\label{ssec:lucy}

To numerically invert the integral equation relating $w(\theta)$ to 
$P(k)$, we must first approximate the integral as a discrete summation.
The power spectrum is binned into $m$ wavenumbers, $k_{j}$, and 
the angular correlation function is tabulated at $n$ angles, $\theta_{i}$. 
For the $r^{\rm th}$ iteration of the inversion algorithm, the estimate 
of the power spectrum, $P^{r}(k_{j})$, is used to generate a corresponding 
estimate of the angular correlation function: 
\begin{equation}
w_i^r=\sum_j g_{ij} \ k_j^2 P_j^r \ \Delta \ln k.
\label{eq:matrix}
\end{equation}
Here, $g_{ij}$ is a discretized version of the kernel, in the form of an 
$n \times m$ matrix. 
A revised estimate of the power spectrum, $P^{r+1}$, is obtained by 
comparing the estimated angular correlation function, $w^r$, with 
the measured correlation function $w^0$:  
\begin{equation}
P^{r+1}(k_j)=P^{r}(k_j)
\frac{\sum_i \frac{w^0(\omega_i)}{w^r(\omega_i)} g(k_j \omega_i) 
					\Delta \ln \omega }
 {\sum_i g(k_j \omega_i) \Delta \ln \omega }.
\label{eq:lucy}
\end{equation}

\subsection{Practical implementation of Lucy's method}
\label{ssec:stopping}

Lucy (1974) demonstrated that the algorithm presented in the previous 
subsection results in an estimate of the unknown quantity, in our case 
the power spectrum, that tends to the maximum likelihood 
solution of the integral equation as more iterations are completed. 
However, if too many iterations are performed, the recovered spectrum 
will not necessarily be smooth, as the solution will attempt 
to reproduce all the features in the data, including noise. 
The problem of how to select an optimum iteration at which to stop 
the inversion was not addressed in Lucy's 1974 paper.
Baugh \& Efstathiou (1993,1994) chose the ``best'' iteration by inspection 
of the closeness of the estimated angular correlation function to the 
measured correlation function, whilst at the same time requiring  the 
corresponding power spectrum to be smooth.
Gazta\~naga \& Baugh (1998, GB98) attempted to automate this process 
by computing a $\chi^2$ value to quantify the level of agreement 
between the correlation function obtained from the estimate of the power 
spectrum using Limber's equation and the measured angular 
correlation function. 
A plot was made of the value of $\chi^2$ against iteration number 
and the estimates of the power spectrum corresponding to 
local minima in $\chi^2$ were examined for smoothness. 
In this paper, we apply the inversion technique to over two hundred 
mock catalogues, so neither of the above methodologies is appropriate. 

To devise an objective,  reliable, and automatic way 
to choose the best iteration, 
we follow a slightly modified version of the approach proposed by Lucy (1994).
Lucy wrote down expressions for quantities related to the 
likelihood and smoothness of the solution to the integral equation. 
The stopping criterion was determined by the way in which the values of 
the smoothness and likelihood functions changed between consecutive iterations.

We have tailored the stopping criteria devised by Lucy to reflect the 
characteristics of angular correlation function data.
In our case, the angular correlation function is positive and is
a power law at small angles. 
The power law has a break around $\theta=2^o$ for the case of the APM survey,
after which the correlation function drops rapidly in amplitude with 
increasing angular separation.
For each iteration, we calculate a $\chi^2$ parameter by comparing
the observed correlation function, $w^0(\theta_i)$, with 
the estimated correlation function, $w^r(\theta_i)$, using $20$ bins 
in $\theta$ around the break, up to the angle for which 
$w^r(\theta_i)=2 \times 10^{-3}$, at bin $i_b$:\footnote{The motivation for 
stopping the comparison once $w(\theta)$ reaches this amplitude is 
that this is the level of the expected error introduced by the procedure used 
to calibrate the magnitude scale between APM plates; see 
Maddox, Efstathiou \& Sutherland (1996).} 
\begin{equation}
\left(\chi^2\right)^r=\sum_{i_b-20}^{i_b} (w^0(\theta_i)-w^r(\theta_i))^2.
\label{eq:good}
\end{equation}
The value of $(\chi^2)^r$ is small if the break
in the observed angular correlation function is well reproduced by the
$r^{th}$ iteration of the inversion. 
We also require a smooth solution for the
power spectrum, in accordance with theoretical expectations, 
given the width of the bins in wavenumber 
used to tabulate $P(k)$. 
We quantify the smoothness of the solution $P^r(k)$, by computing 
the following quantity: 
\begin{equation}
C^r= \sum_{i=1}^{n} (P^r(k_i) - P^r(k_{i-1}))^2;
\end{equation}
smoother solutions for $P(k)$ will result in 
smaller values of the function $C^r$. 

We normalise $(\chi^2)^r$ and $C^r$ to their peak values over 40 
iterations, and define the goodness of the
$r^{th}$ iteration by the function: 
\begin{equation}
G^r= {\rm abs}[(\chi^2)^r]^{\alpha} + {\rm abs}(C^r)^{\beta},
\label{eq:goodness}
\end{equation}
where the exponents are set to $\alpha=0.5$ and $\beta=1.0$, 
reflecting the relative weights given to reproducing the 
angular correlation data (in the case of $\alpha$) or the 
smoothness of the derived solution (in the case of $\beta$).
The first local minimum in the quantity $G^r$ is adopted as the 
best solution to the integral equation. The precise location of 
the first local minimum is insensitive to the exact choice of 
values for $\alpha$ and $\beta$. 

\subsection{Evolution of the power spectrum}
\label{ssec:ev}
\begin{figure}
{\epsfxsize=8.truecm 
\epsfbox[50 180 320 695]{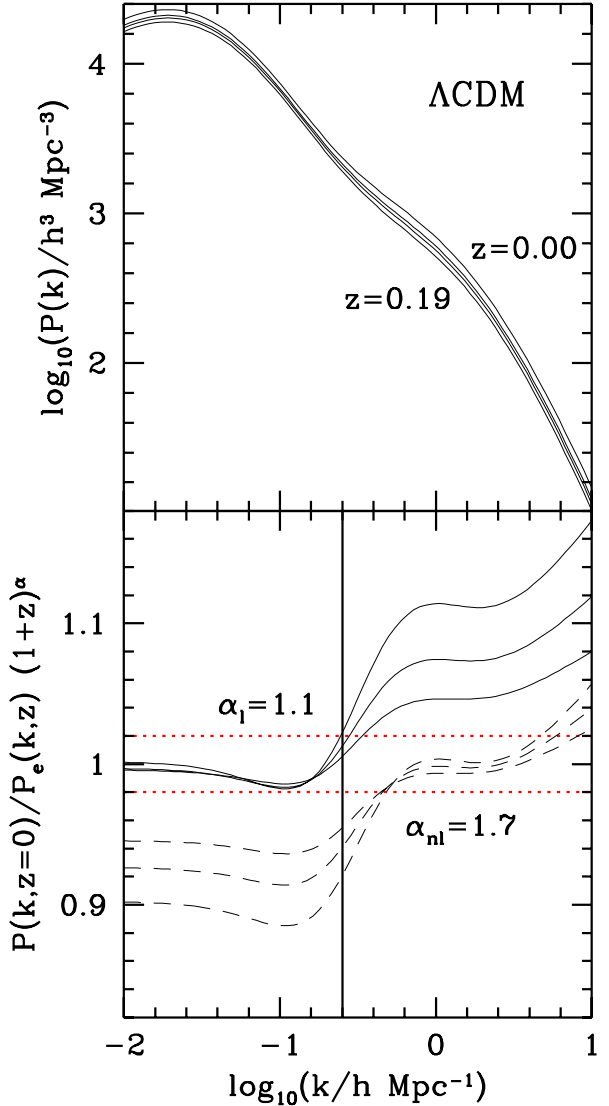}}
\caption{
Redshift evolution of the power spectrum of density fluctuations in 
a flat, $\Lambda$CDM model ($\Gamma=0.2$, $\sigma_{8}=0.9$, $\Omega_{m}=0.3$).
The upper panel shows the nonlinear power spectrum at 
redshifts $z=0.0, 0.09, 0.13,$ and $0.19$ (from top to bottom in amplitude).  
The lower panel shows the ratio $P(k,z=0)/P(k,z) (1+z)^{\alpha}$, 
for $\alpha=1.1$ (solid lines) and $\alpha=1.7$ (dashed lines)
for the different redshifts.  The vertical line shows the
transition scale where the density fluctuations
become nonlinear according to the condition $\sigma(k)=0.1$, as 
explained in the text.
The horizontal dotted lines delineate the range in which the ratio of 
power spectra falls within $2\%$ of unity.
}
\label{fig:pkevol}
\end{figure}

The value of the parameter $\alpha$ that describes the evolution
of the power spectrum with redshift (see equation \ref{eq:kernel})
depends upon the data under consideration.
In the case of the mock surveys, for example, we use $\alpha=0$, 
since these catalogues are extracted from the $z=0$ output from 
an N-body simulation, so by construction there is no evolution of 
clustering within these samples.

The choice of an appropriate value of $\alpha$ to describe the 
evolution of clustering of galaxies in the APM Survey is less 
straight forward. The problem can be broken down into 
two components: (i) quantifying the evolution of clustering in 
the underlying dark matter distribution, and (ii) modelling 
the redshift dependence of galaxy bias.
Semi-analytic models of galaxy formation predict relatively 
little evolution in galaxy bias for flux limited samples over 
the redshift interval spanned by the APM Survey (Baugh \etal 1999), 
so we will ignore the latter of these effects.
We therefore focus our attention on the evolution of the matter 
power spectrum. 

We compute the nonlinear power spectrum of matter fluctuations using 
the transformation proposed by Smith \etal (2002), which describes 
the results of N-body simulations to an accuracy of $9\%$.
In the upper panel of Fig. \ref{fig:pkevol}, we plot the power spectrum 
calculated at $z=0$ and at $z=0.09,0.13$ and $0.19$. The latter three  
values correspond, respectively, to the redshifts at which 
we expect to see $25\%$, $50\%$ and $75\%$ of the galaxies in 
the APM Survey, according to our model for the redshift 
distribution (see \S \ref{ssec:nofz}).
We use the Efstathiou, Bond \& White (1992) fit to the linear theory 
cold dark matter power spectrum with shape parameter, $\Gamma =0.2$, 
with a present day amplitude, 
specified in terms of the variance in spheres of radius $8h^{-1}$Mpc, 
of $\sigma_8=0.9$.  
We assume a flat universe with a present day density parameter 
of $\Omega_{m}=0.3$ and a cosmological constant.

In the lower panel of Fig. \ref{fig:pkevol}, we show the ratio  
of the present day nonlinear power spectrum to the power spectra 
computed at redshifts $z=0.09,0.13$ and $0.19$, including the correction for 
evolution given by equation (\ref{eq:alpha}). 
There are two sets of curves corresponding to two different 
choices for the value of $\alpha$.  
The solid vertical line gives a rough indication of the wavenumber at which we 
expect the density fluctuations to become nonlinear. This is estimated 
by computing the variance of counts-in-cells of APM galaxies as a function 
of scale, and adopting $\sigma = 0.1$ as a criteria for the onset of 
nonlinear behaviour (see e.g. Baugh \& Efstathiou 1994b; note 
that we have also 
assumed that the bias between fluctuations in galaxies and the underlying 
mass is unity).
In practise, the approximation for the evolution of clustering given 
by equation (\ref{eq:alpha}) is accurate to within $\sim 15\%$ over the 
range of scales we are interested in. Over a more restricted range of 
scales, the approximation performs much better. For wavenumbers on which 
we expect the fluctuations to still be in the linear regime, 
$\alpha = 1.1$ describes the evolution of the power spectrum to 
an accuracy of $2\%$. On nonlinear scales (as quantified $\sigma(k)>0.1$), 
a choice of $\alpha=1.7$ gives an excellent description of the 
evolution of $P(k)$.
We will therefore use both of these values of $\alpha$ in the 
inversion of the APM angular clustering data.

\subsection{Mock catalogues}
\label{ssec:mocks}

Mock catalogues constructed from the Virgo Consortium's
Hubble Volume Simulations (Evrard \etal 2002) are a key 
component of the inversion algorithm presented in this paper. 
Full details of the construction of these mock catalogues and 
their properties are given by Baugh et al. (2003, in preparation, B03; 
the fabrication of the mock catalogues is also described in 
Norberg \etal 2002a).
Here we reiterate a few of the points pertinent to our analysis.

The mock catalogues should, by design, display the same clustering as 
APM Survey galaxies. The clustering of the dark matter in the  
Hubble Volume simulations is different from that measured 
for APM Survey galaxies, particularly in the case of $\tau$CDM 
(Jenkins \etal 1998). 
Therefore it is necessary to choose particles 
from the simulation in such a way that the clustering signal of 
the dark matter at the present day is modulated to match 
the required galaxy clustering.  
We use the second biasing algorithm from Cole \etal (1998), in 
which the probability of selecting a dark matter particle to be 
a ``galaxy'' is a function of the smoothed dark matter density 
at the present day.

The choice of ``observers'' around which to build the mock catalogues 
is based on two criteria.
The first of these places the observer at a location in the 
simulation around which the characteristics of the density 
and velocity fields are similar to those measured for the 
local group (denoted local group observer mocks, see B03 for full details 
of the conditions).
The largest volume constraint used to define a local group environment 
corresponds to a radius of $70h^{-1}$Mpc. In the $\Lambda$CDM 
Hubble Volume simulation, 5000 independent test observers could be packed 
into the simulation volume; out of these, only 22 satisfied the local 
group selection criteria.
The second criterion simply places observers at 
random locations within the simulation box (hereafter referred to as random 
observer mocks), with the condition that the observers must be separated 
by a distance of at least $700h^{-1}$Mpc, in order to minimise overlap 
between catalogues.
We use 44 local group observers and 21 random observers in total about 
which to extract mocks from the two Hubble Volume simulations.

Once an observer has been positioned within the Hubble Volume box, ``galaxy'' 
particles are chosen in such a way as to reproduce the expected 
radial selection function of APM galaxies.  
We use a model for the APM selection function based on results 
from the 2dFGRS (see Section \ref{ssec:nofz} and Norberg et al. 2002a). 
The catalogues extend to a magnitude limit of $b_{\rm J}=20$; 
we explain the extension of the 2dFGRS selection function to 
this depth in Section \ref{ssec:nofz}. 
The mock catalogues possess the same angular mask as the
APM Galaxy Survey (Maddox \etal 1990). 
The median redshift of the mock catalogues is $z_m \simeq 0.115$, 
and they each cover a solid angle of $1.3$ steradians. 
In most subsequent analyses, the mock catalogues are divided 
into four zones, equally spaced in right ascension, with a width of around 
30 degrees (see Fig. 2 of Baugh \& Efstathiou 1994a).

In addition to testing the inversion algorithm, we use mock catalogues 
to obtain a reliable estimate of the uncertainties in the inversion 
results, which is an important advance over much previous work.
This requires that we make the assumption that the mock catalogues 
are independent. 
Strictly speaking, this is incorrect because we have used simulations that 
are single realisations of large volumes. Long wavelength fluctuations 
can run through the whole of the simulation box, introducing coherency 
in the density field sampled by different observers. 
In practise, this is not an issue as such perturbations are incredibly 
weak, given the form of the CDM power spectrum.
Even in the case of local group observers, where one might 
worry that the sphere of exclusion around each observer is 
comparatively small, the relative orientation of the mock 
catalogues and the influence of the selection function mean 
that there are unlikely to be significant numbers of particles 
found in common in the mocks around adjacent observers. 
The only way to have formally independent mocks is to run a 
dedicated N-body simulation for each catalogue. This is computationally 
prohibitive if one wishes to retain the accurate modelling of long 
wavelength perturbations conferred by using a box size comparable to 
that employed in the Hubble Volume simulations.

\subsection{Redshift distribution}
\label{ssec:nofz}

\begin{figure}
{\epsfxsize=8.truecm 
\epsfbox[50 178 570 696]{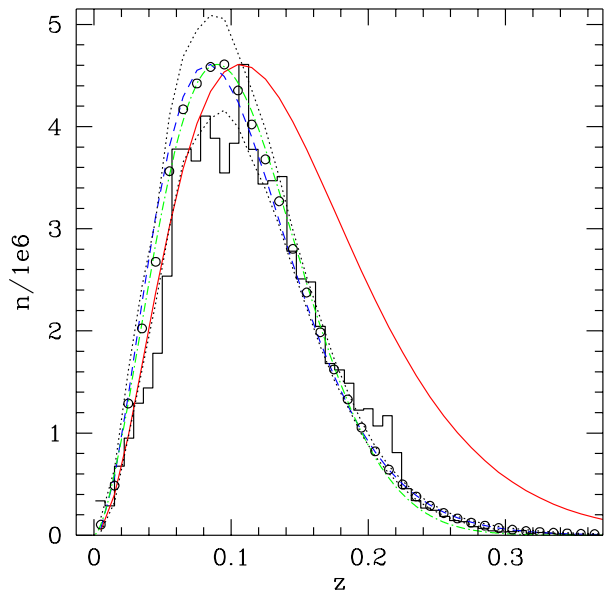}}
\caption{
Redshift distribution of the 2dFGRS 100K release
(histogram) and the theoretical $n(z)$ obtained by integrating 
the luminosity function (long-dashed line).
The mean $N(z)$ measured from the mock catalogues, magnitude limited to 
$b_{\rm J}=19.45$, is shown by the open circles;
the dotted lines show the $68\%$ confidence range in $n(z)$ 
obtained from the mock catalogues.
The parametric fit given by equation (\ref{eq:nzfit}) is shown by 
the short dashed line. The solid line gives the same fit evaluated for 
a magnitude limit of $b_{\rm J}=20$.
}
\label{fig:nz_2df_mocks}
\end{figure}

\begin{figure}
{\epsfxsize=8.truecm 
\epsfbox[20 178 570 696]{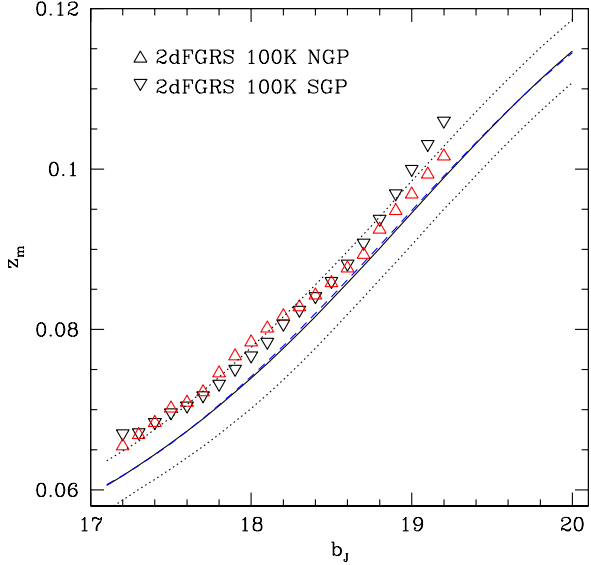}}
\caption{
Median redshifts for the 2dFGRS as a function of faint apparent
magnitude limit. The triangles show the results from
galaxies in the NGP (upward pointing triangles) and SGP galaxies 
(downward pointing triangles). The solid line is the result of
integrating the 2dFGRS luminosity function, and the dotted lines show
the $90\%$ confidence levels as obtained from mock catalogues. The
dashed line corresponds to the result obtained from
a simple parametric fit to the redshift
distribution, given by equations (\ref{eq:nzfit}) and (\ref{eq:zm}).
}
\label{fig:zm}
\end{figure}

The selection function used to construct the mock catalogues is 
determined by the $z=0$ luminosity function estimated from the 2dFGRS 
and by the form adopted for the $k+e$ correction (Norberg \etal 2002a).
At present, the 2dFGRS luminosity function estimate is the most accurate 
available, displaying, for the first time, random measurement errors that 
are smaller than the systematic errors (e.g. choice of $k+e$ correction, 
uncertainty in the zero-point of the magnitude scale; see Norberg \etal 2002a 
for a detailed discussion of such effects).
The $k+e$ corrections adopted by Norberg \etal (2002a), which compensate for 
the band-shifting of the $b_{\rm J}$ filter with redshift and for changes 
in the stellar population of the galaxy, were first proposed by Zucca \etal 
(1997), based on stellar population synthesis models. 
A key test of these corrections was carried out by Norberg et al. These 
authors found that the luminosity function estimates from two subsamples 
drawn from the 2dFGRS covering disjoint redshift intervals were in 
excellent agreement. This means that the form adopted for the $k+e$ 
correction provides an accurate description of the evolution of 
galaxy luminosity, at least over the redshift range probed 
in the test. 
In the construction of the mock catalogues, B03 retain the form for the 
$k+e$ correction used by Norberg et al., pushing the model to 
a fainter magnitude limit and, therefore, to a slightly higher 
redshift than it has been tested.

The inversion algorithm requires the redshift 
distribution of galaxies to be specified.
The redshift distribution can be estimated by integrating over the 
luminosity function and volume element as follows: 
\begin{equation}
n(z) = \int_0^{\infty} {\rm d}z \frac{{\rm d}V}{{\rm d}z}
 \int_{M_1(z)}^{M_2(z)} \Phi(M){\rm d}M,
\label{eq:nzlf}
\end{equation}
where $$M(z=0)=m- \ [k+e](z) \ -5 \log(d_L/h^{-1} {\rm Mpc})-25,$$ 
and $m$ is the apparent magnitude limit of the survey 
and the mean $k+e$ correction as a function of redshift is given by 
$k+e(z)=(z+6z^2)/(1+20z^3)$ (Norberg et al. 2002a).
The redshift distribution obtained by applying equation (\ref{eq:nzlf})
to the 2dFGRS apparent magnitude limit of $b_{\rm J}=19.45$ is shown 
by the long-dashed line in Fig. \ref{fig:nz_2df_mocks}. 
The redshift distribution from the 2dFGRS 100K data release is shown 
by the histogram.
At first sight, the theoretical redshift distribution obtained by 
integrating over the luminosity function does not appear to be the 
best fit to the histogram of observed redshifts. This is due 
to large scale structure, which we demonstrate by plotting the average 
$n(z)$ from the mock catalogues, magnitude limited to $b_{\rm J}=19.45$ 
(shown by the open circles). There is a significant spread in the $n(z)$ 
curves from mock-to-mock, illustrated by the dotted lines, which show 
the range containing $68\%$ of the mocks. 
The mean $n(z)$ from the mocks, averaging over 
fluctuations arising from large-scale structure is in excellent agreement 
with the prediction from equation (\ref{eq:nzlf}, see also the discussion 
in Norberg \etal 2002a, in particular for possible explanations of the feature 
seen in the data at $z \sim 0.2$).

The average redshift distribution obtained from the $65$ mock 
APM catalogues is accurately described by the parametric form: 
\begin{equation}
\left( \frac{{\rm d}N}{{\rm d}z}\right) {\rm d}z = A z^2 \exp \left(
-(z/z_c)^{\beta}
\right) {\rm d}z,
\label{eq:nzfit}
\end{equation}
with parameters $z_c=0.072$ and $\beta=1.23$ for $17 < b_{\rm J} < 20$.
In this expression, $z_c$ is related to the median redshift of the survey, $z_m$, through 
\begin{equation}
z_c  =  z_{m}/1.82 = a\left[ (b_{\rm J}-16)^2 \exp(-b (b_{\rm J}-16)^3) +c \right], 
\label{eq:zm}
\end{equation}
where $a=2.8 \times 10^{-3}$, $b=4.8 \times 10^{-3}$, and $c=10.7$. 
The fit to the 2dFGRS data given by equation (\ref{eq:nzfit}) is shown 
by the dashed line in Fig. \ref{fig:nz_2df_mocks}. 
The $n(z)$ for galaxies magnitude limited to $b_{\rm J}=20$ is given 
by the solid line in Fig. \ref{fig:nz_2df_mocks}. 

A further test of our parametric fit to the 2dFGRS redshift distribution 
is to compute the median redshift, $z_{m}$, as a function of apparent 
magnitude limit.
In figure \ref{fig:zm}, we show this estimate of $z_m$ by the solid line. 
The median redshift for the 2dFGRS 100K data release is shown by 
the triangles (upward pointing triangles correspond to the NGP region
and downward pointing triangles to the SGP). 
The uncertainty in the estimate of the median redshift is derived by 
obtaining $z_m$ for each of the mock catalogues 
and determining the redshift range that contains $90 \%$ of the mocks 
(shown by the dotted lines). 
The median redshifts measured in the two 2dFGRS regions lie near the upper 
$90 \%$ confidence level line. It should be noted though, that
the solid angle of the mock catalogues corresponds to that of the
APM catalogue, which is roughly a factor of $4$ larger than the solid
angle of the combined SGP and NGP 2dFGRS samples.
This means that the errors plotted in this figure
should be considered as a lower limit to the errors we would 
expect for the 2dFGRS solid angle.
The dashed line corresponds to the simple analytical fit given in eq.
(\ref{eq:nzfit}), which is in excellent agreement with the result obtained 
from the theoretical prediction for $n(z)$ (i.e. derived by integrating over 
the luminosity function).

\subsection{Error estimation}
\label{ssec:error_estimation}

The primary role of the mock catalogues is in the estimation of the 
errors on the recovered power spectrum or on the parameters of theoretical 
models for the power spectrum. 
The only approximation in our approach is that the mock catalogues 
provide a realistic description of the large-scale structure of 
the Universe.

We use the scatter in the recovered power spectra from 260 mock 
APM zones (65 mock observers, with each catalogue split into four zones) 
to infer the errors. 
The advantage of using such a large number of mocks is that we can take 
into account any possible covariance between the power spectrum measured 
at different wavenumbers.  
Correlations can be introduced by the non-linear growth of the density 
field (Meiksin \& White, 1999) or by the finite width of the 
kernel function in eq. (\ref{eq:kernel}).

We experimented with computing a formal covariance matrix using 
the mock catalogues. We found that even with 260 mocks, the 
off-diagonal components of the covariance matrix are noisy and 
cannot be obtained reliably.
Instead, we use the {\it rms} scatter between the results obtained 
for the individual mocks to compute the errors, which, at some level, 
indirectly takes into account the correlations between measurements 
on different scales. This approach was followed by Norberg \etal 
(2001,2002b) when estimating the errors on the two-point galaxy 
correlation function from the 2dFGRS.
Two types of result are presented in this paper and we outline 
how the errors are estimated on each below.

\noindent
(i) The power spectrum of APM Survey galaxies as a function of wavenumber. 
The {\it rms} scatter in the power is built up by treating the 
power spectrum recovered for each mock APM zone in turn as the 
``reference spectrum'' and computing the variance with respect to this 
using the measurements in the remaining mocks.
We effectively compute the mean of 260 estimates of the variance in 
this way.   

\noindent
(ii) The cosmological parameters that specify the cold dark matter 
power spectrum.
The best fitting parameters are determined for each mock by 
minimising $\chi^{2}$, using the remaining mocks to set the variance 
on the power spectrum. This process is repeated for the full ensemble 
of mocks and the errors on the best fitting cosmological parameters 
are taken to be the resulting {\it rms} scatter.

In some parts of the paper, we quote the error on the 
mean of our measurements. As the real APM Survey is divided 
into four zones, we have four measurements which we take to 
be independent, so we divide the {\it rms} scatter by $\sqrt{3}$ 
to obtain the error on the mean from the variance.

\subsection{Testing the inversion technique on mock catalogues}
\label{ssec:app_mocks}

\begin{figure}
{\epsfxsize=8.truecm 
\epsfbox[20 146 568 696]{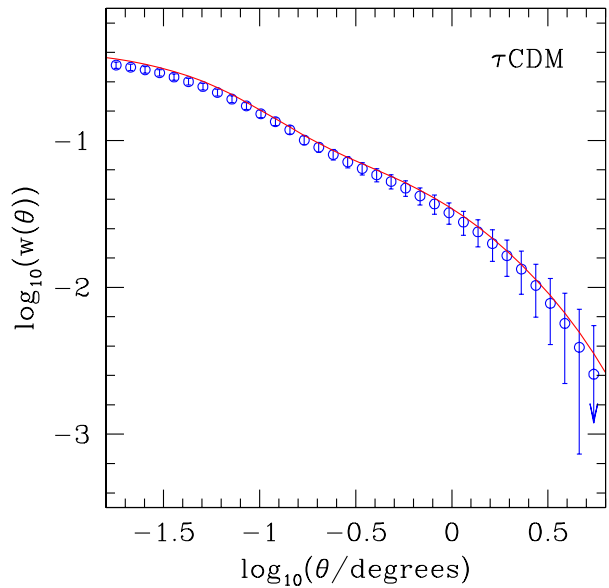}}
\caption{
The points show the mean $w(\theta)$ measured directly in the $\tau$CDM 
APM mock catalogues. The errorbars show the {\it rms} scatter over the 
ensemble of mocks. 
The solid line shows an estimate of $w(\theta)$ computed by applying  
eqn. (\ref{eq:limber}) to the power spectrum obtained from an FFT of the 
distribution of biased ``galaxy'' particles in the full simulation volume.
}
\label{fig:wbr}
\end{figure}

\begin{figure}
{\epsfxsize=8.truecm 
\epsfbox[20 146 568 696]{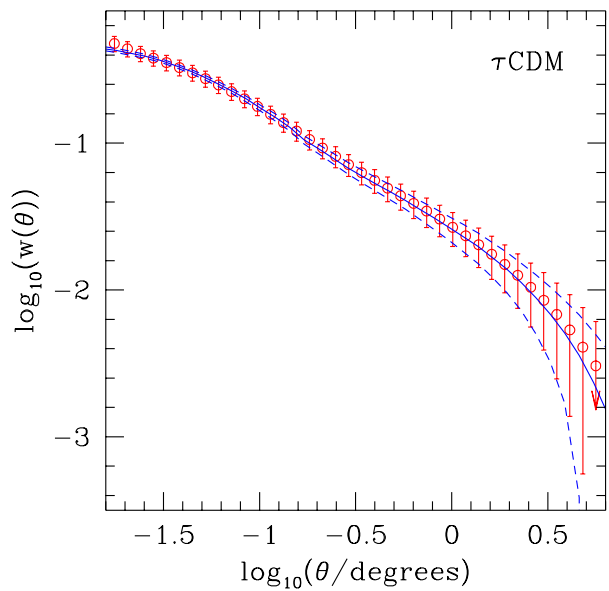}}
\caption{
The points show the mean angular correlation function computed from 
the best estimate of the power spectrum in each mock APM zone. 
The errorbars show the {\it rms} scatter. The solid line shows the 
mean of the direct measurements of $w(\theta)$; the dashed lines connect 
the tops and bottoms of the errorbars showing the {\it rms} scatter on 
the direct measurements.
}
\label{fig:wbl}
\end{figure}


\begin{figure}
{\epsfxsize=8.truecm 
\epsfbox[46 250 322 679]{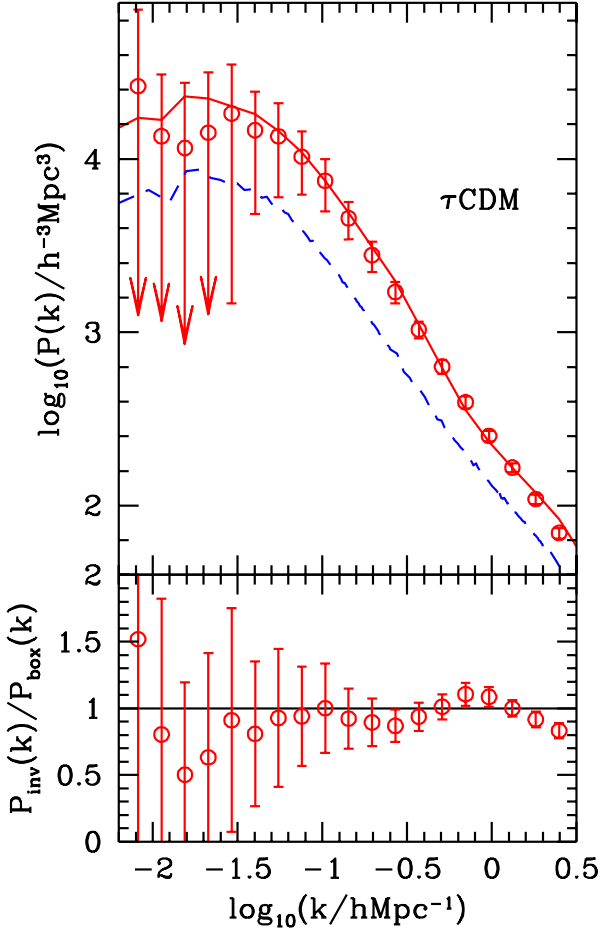}}
\caption{
The upper panel shows the mean power spectrum recovered by inverting 
the angular correlation function in the $120$ zones in the APM mock 
catalogues (open circles) drawn from the $\tau$CDM simulation.  
The solid line shows the power spectrum measured directly by applying 
an FFT to the distribution of biased ``galaxy'' particles in the 
full simulation box; the dashed line shows the power spectrum of the
dark matter particles in the simulation.
The ratio of the mean inverted $P(k)$ 
to the FFT power spectrum is shown in the lower panel (open circles). 
The errorbars show the {\it rms} scatter in this ratio, and 
the solid line corresponds to a ratio of unity.
}
\label{fig:lucy_av}
\end{figure}

In this Section, we establish the accuracy of the inversion technique 
and the error estimation procedure using mock catalogues constructed from the 
Hubble Volume simulations. 
To save space, we show plots only for the $\tau$CDM Hubble Volume mocks; 
the $\tau$CDM case provides a more convincing test of the inversion method 
as the clustering signal imposed on the mock catalogues is very different 
from the clustering of the underlying dark matter (see figure
\ref{fig:lucy_av}).

We compare the power spectrum recovered by inverting the angular correlation 
estimated in mock APM zones with the power spectrum measured from the full 
simulation box. To properly interpret this comparison, it is necessary to 
check if there are any systematic differences between the clustering 
measured in an ensemble of mock catalogues and that measured from the 
full simulation volume. This test is presented in Fig. \ref{fig:wbr}. 
The solid line shows an estimate of the angular correlation function 
obtained by applying eqn. (\ref{eq:limber}) to the power spectrum 
measured by a direct fast Fourier transform of the distribution of 
biased ``galaxy'' particles in the entire simulation box.  The regriding 
technique devised by Jenkins \etal (1998) was used so that the result 
of the FFT is not influenced by aliasing due to the grid assignment scheme.
The points show the mean angular correlation function obtained 
by averaging over the measurements from the mock catalogues. 
The error bars show the variance over the ensemble of mocks. The volume 
probed by the mock APM zones is sufficient to yield a measurement of 
the two-point correlation function that agrees with the full box result 
to better than $10\%$. 

In order to apply the inversion algorithm described in Section 
\ref{ssec:lucy}, we use a fit to the mean redshift distribution 
found in the mock catalogues, utilising the form given in 
equation (\ref{eq:nzfit}). 
The best iteration for each mock is determined by searching for 
local minima in the ``goodness'' measure, $G$, defined by equation 
(\ref{eq:goodness}).

A first illustration of the accuracy of the inversion is given 
by Fig. \ref{fig:wbl}, in which we compare the mean angular correlation 
function obtained by averaging the $w(\theta)$ corresponding 
to the best estimate of the power spectrum in each zone (points) to 
the mean of the $w(\theta)$ measured directly in each zone (solid line). 
The errorbars show the variance in the $w(\theta)$ obtained from the 
recovered power spectra, and the dashed lines connect the tops and 
bottoms of the errorbars that indicate the variance on the direct 
measurements of $w(\theta)$ from the mocks. 
The mean $w(\theta)$ computed from the inversion results is in excellent 
agreement with the mean of the direct measurements; the scatter in 
$w(\theta)$ from the recovered $P(k)$ is in good agreement with the 
intrinsic scatter in the mocks, demonstrating that our criteria for 
stopping the iterative scheme works well.

The mean of the recovered power spectrum is shown by the points in 
upper panel of Fig. \ref{fig:lucy_av}. The errorbars show 
the {\it rms} scatter over the ensemble of mock catalogues, 
computed as described in Section \ref{ssec:error_estimation}. The solid 
line shows the power spectrum measured for the biased particles 
in the full simulation volume, using the FFT technique described 
above.   The dashed line shows this result for the dark matter particles
in the simulation.
The ratio between the recovered spectrum and the FFT estimate 
is plotted in the lower panel. The errorbars show the {\it rms} scatter 
from the mocks. 
In view of the slight offset between the angular clustering 
obtained by averaging over the mock catalogues and that inferred 
from the power spectrum measured from the full simulation 
box (as shown in Fig. \ref{fig:wbr}), 
one would expect the inversion results to slightly underestimate the 
power spectrum measured from the full volume. This is indeed the 
case, although the ratio of power spectra is somewhat noisier than it is 
for the angular correlation function. The inversion algorithm 
recovers the power spectrum to within $10-20$\% over most of the 
range of wavenumbers plotted. Moreover, given the size of the errorbars 
that we have estimated using the mock catalogues, the inversion results 
are in excellent agreement with the power spectrum measured using an FFT, 
over almost two and a half orders of magnitude in wavenumber.

\subsection{Comparison of errors with other estimates}
\label{ssec:error_comp}
\begin{figure}
{\epsfxsize=8.truecm 
\epsfbox[20 170 580 700]{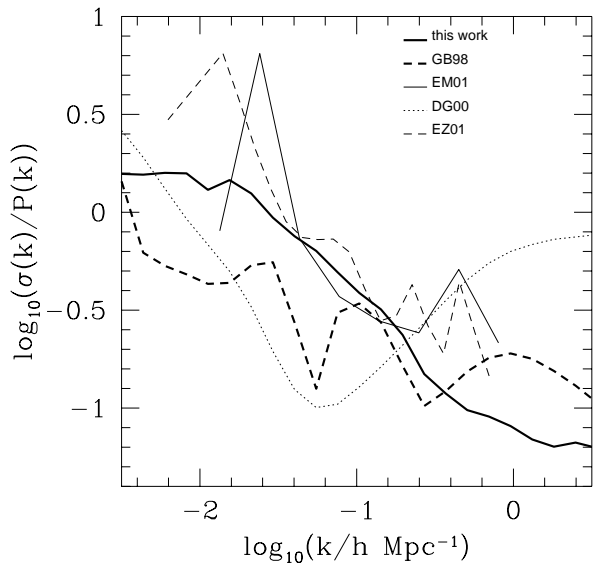}}
\caption{
The fractional variance in the power spectrum as a function of wavenumber 
obtained in this work from mock APM catalogues (heavy solid line). 
For comparison, we also show the estimates of Efstathiou \& Moody 
(2001, thin solid line), EZ01 (thin dashed line), and DG00 (dotted line).  
We also show a previous estimate obtained by 
Gazta\~naga \& Baugh (1998, thick dashed line), rescaled as described in the 
text.
}
\label{fig:errors}
\end{figure}

\begin{figure}
{\epsfxsize=8.truecm 
\epsfbox[20 170 580 700]{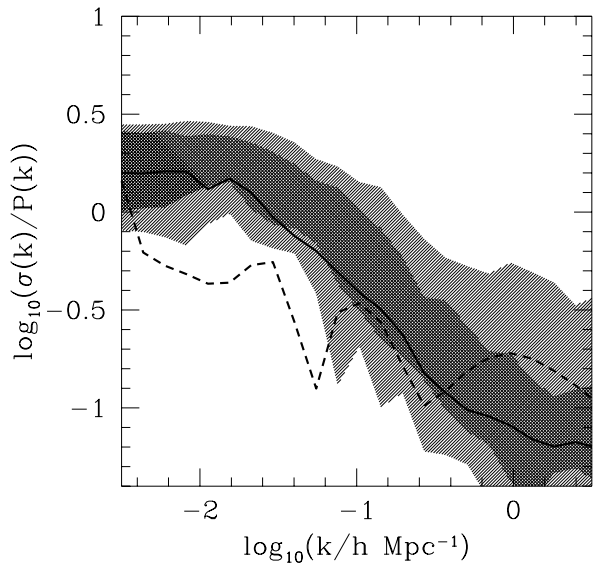}}
\caption{
The fractional variance in the power spectrum as a function 
of wavenumber obtained in this work from mock APM catalogues 
(thick solid line) and a previous estimate by GB98 (heavy dashed line).
The shaded areas show the range of variances obtained from the scatter 
between the four zones that make up each mock APM catalogue. 
The heavy shading encloses $68$\% of the mock catalogues and 
the light shading extends this region to encompass $95$\% of the mocks.
}
\label{fig:errors_mocks}
\end{figure}

It is instructive to compare the error estimates presented 
in this paper with those published in the literature, which 
generally rely upon a different set of assumptions and approximations.
Essentially the only assumptions we make are that the mock 
catalogues give a realistic picture of the large-scale structure 
of the Universe and that our approach of computing the {\it rms} 
scatter over the ensemble of mocks takes into account, at some 
level, the covariance between measurements on different 
scales.  
In this Section, we show results from the $\tau$CDM and $\Lambda$CDM 
mocks in combination.

The fractional error in the power spectrum, $\sigma(k)/P(k)$, is plotted 
in Fig. \ref{fig:errors}. The thick solid line shows the {\it rms} scatter 
over the full ensemble of mocks. 
The thin solid line shows the error estimate of Efstathiou \& Moody (2001), 
who used a maximum likelihood approach to estimate $P(k)$, and assumed 
a Gaussian distribution of density fluctuations. 
The light dashed line shows the estimate of Eisenstein \& Zaldarriaga (2001); 
these authors  performed a SVD matrix inversion of Limber's equation, assuming 
a Gaussian model for the error in $w(\theta)$ at large angles. 
The dotted line shows the error estimate of Dodelson \& Gazta\~{n}aga (2000). 
DG00 again assume Gaussian errors in $w(\theta)$ and require a value to be 
chosen for a smoothing parameter, which has some influence on the errors.
Overall, the error estimates of Efstathiou \& Moody and Eisenstein \& 
Zaldarriaga are in good agreement with our estimate, on the scales for 
which the assumption of Gaussianity makes a comparison possible.

The heavy dashed line in Fig. \ref{fig:errors} shows the error 
estimate from GB98. The errors quoted in Table 2 of GB98 are the 
error on the mean and not the variance. These authors split 
the APM survey into four zones to invert the angular correlation 
function, and obtained four estimates of the power spectrum, which 
they treated as independent. Therefore they divided the scatter 
in $P(k)$ between zones by an additional factor of $\sqrt{3}$ to get the 
error on the mean. 
Note that the GB98 error estimate neglects any 
correlation between wavenumber bins.
In order to compare GB98's estimate of the errors with the 
fractional variances plotted in Fig. \ref{fig:errors}, we have therefore 
multiplied their estimates by $\sqrt{3}$.
At larger wavenumbers, $\log(k/h{\rm Mpc}^{-1})>-1$, the GB98 estimate 
of the errors agrees with our estimate. On larger scales, GB98 underestimate 
the errors by up to a factor of two, as noted by EZ00 and EM01.
We can use the mock catalogues to repeat the error analysis of GB98, 
using the scatter between the zones in each mock catalogue to make an 
estimate of the errors on the recovered power spectrum. 
In Fig. \ref{fig:errors_mocks}, the dashed and 
solid lines are the same as those plotted in Fig. \ref{fig:errors}. 
The shaded regions show the range of variances obtained by 
taking each full, mock APM catalogue, consisting 
of four zones, and using just these zones to estimate the variance. 
The darker shading encloses the range in which $68$\% of the zone variances 
from the catalogues fall and the light shading shows the $95$\% range. 
The conclusion reached upon examination of this figure, is, that on 
large scales, GB98 were simply unlucky in their estimate of the 
errors. The scatter between the power spectra recovered in the zones 
of a given mock catalogue is only as small as that estimated by GB98 from 
the APM Survey zones in fewer than $5\%$ of the mocks.

These results clearly show the benefit of using mock catalogues in 
the estimation of confidence levels for statistics, such as the 
power spectrum, which may suffer from bin-to-bin correlations. The mocks 
can be used to obtain a reliable estimate of the error on the 
power spectrum over a wider range of scales than is possible if Gaussian 
fluctuations are assumed.   
In particular, our method allows us to extend
the measurement of the power spectrum and the estimate of 
the uncertainty on this measurement into the small-scale, 
high wavenumber region, which is powerful for constraining 
models of galaxy formation. 
We have checked that the errors from the random observer mocks
are consistent with those derived from the local group observer mocks;
the errors differ at most by $25 \%$ in amplitude. 

\section{Application of the method to the APM Galaxy Survey}
\label{sec:app}

\begin{figure}
{\epsfxsize=8.truecm 
\epsfbox[20 170 400 670]{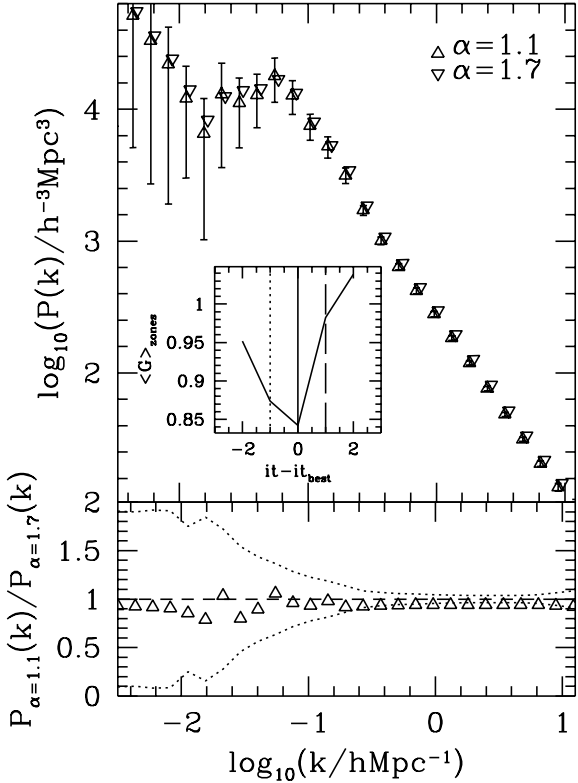}}
\caption{
The upper panel shows the APM power spectrum recovered for two different 
choices of the clustering evolution parameter $\alpha$; the upwards pointing
triangles show the results for $\alpha=1.1$, and the downwards pointing 
triangles those for $\alpha=1.7$.  The lower panel shows the ratio between
these two estimates (open triangles).  The dashed line
shows a ratio of unity and the dotted lines show the fractional error 
on the ratio, as inferred from the {\it rms} scatter from the mock 
catalogues. The inset in the upper panel shows the goodness function, $G$, 
used to identify the best iteration.
}
\label{fig:apm_alpha}
\end{figure}

\begin{table}
\caption{\small
{The power spectrum of APM Survey galaxies. The first 
column gives the wavenumber, $k$, the second and third columns give the 
mean power spectrum averaging over the four zones of the APM Survey for 
two different choices of the value of the parameter $\alpha$. The final 
column gives the fractional {\it rms} scatter obtained from the mock 
catalogues. The absolute error is obtained by multiplying the fractional 
error by the mean power spectrum.
}}
\begin{tabular}{cccc}
\hline
\hline
\noalign{\vglue 0.2em}
$k/h$ Mpc$^{-1}$ & $P(k)/h^3$ Mpc$^{-3}$ & $P(k)/h^3$ Mpc$^{-3}$ & relative \\
 		 & $(\alpha=1.1)$        & $(\alpha=1.7)$        & errors\\
\noalign{\vglue 0.2em}
\hline
\noalign{\vglue 0.2em}
0.0032 &      80260 &       86500 &   0.91\\
0.0043 &      51170 &       55350 &   0.90\\
0.0060 &      33150 &       36100 &   0.92\\
0.0082 &      21910 &       24190 &   0.91\\
 0.011 &      12120 &       14160 &   0.75\\
 0.015 &       6537 &        8296 &   0.84\\
 0.021 &      12960 &       12550 &   0.72\\
 0.029 &      11160 &       13930 &   0.54\\
 0.040 &      12860 &       14400 &   0.44\\
 0.055 &      17880 &       16910 &   0.37\\
 0.076 &      12790 &       13330 &   0.28\\
  0.10 &       7488 &        8009 &   0.22\\
  0.14 &       5226 &        5340 &   0.18\\
  0.20 &       3156 &        3445 &   0.14\\
  0.27 &       1718 &        1861 &   0.086\\
  0.37 &       1007 &        1080 &   0.068\\
  0.51 &      637.5 &       679.7 &   0.056\\
  0.70 &      420.3 &       447.3 &   0.052\\
  0.96 &      279.3 &       297.2 &   0.047\\
  1.32 &      184.2 &       196.1 &   0.040\\
  1.81 &      119.3 &       127.0 &   0.037\\
  2.49 &      76.07 &       80.99 &   0.038\\
  3.42 &      48.54 &       51.63 &   0.036\\
  4.70 &      31.38 &       33.35 &   0.036\\
  6.46 &      20.50 &       21.82 &   0.046\\
  8.88 &      13.63 &       14.59 &   0.065\\
 12.20 &      10.12 &       10.97 &   0.086\\
 16.75 &      1.772 &       1.334 &   0.103\\
 23.02 &      2.672 &       2.807 &   0.100\\
 31.62 &      1.846 &       1.977 &   0.088\\
\noalign{\vglue 0.2em}
\hline
\hline
\end{tabular}\label{table_pow}
\end{table}

The APM Galaxy Survey is still one of the largest machine constructed 
catalogues of angular positions of galaxies available today 
(Maddox \etal 1990, 1996). It is the parent catalogue of the 2dFGRS, 
but extends over a magnitude deeper than 
the spectroscopic sample; together with the substantially 
larger solid angle covered, this means that the APM Survey covers 
a significantly bigger volume of the local Universe than the 2dFGRS. 
Here we apply the inversion algorithm described and tested in 
Section \ref{sec:inversion} to the angular correlation function 
measured for galaxies brighter than $b_{\rm J}=20$ from the 
original APM Survey area in the SGP. 
This  region was chosen to cover part of the sky that is relatively 
free from obscuration by dust in our Galaxy (see the detailed 
discussion of the effects of Galactic extinction in Efstathiou 
\& Moody 2001).

Following previous studies, we split the survey into four 
zones in the right ascension direction (see Fig. 2 of Baugh \& Efstathiou 
1994a) and apply the inversion algorithm to each zone in turn. 
We perform the inversion for two values of the parameter $\alpha$ that 
describes the evolution of clustering, as justified in Section \ref{ssec:ev}.
The power spectra obtained by averaging the inversion results 
for each zone are shown by the points in the upper panel of 
Fig. \ref{fig:apm_alpha};  the upwards pointing triangles show the 
power spectrum obtained when $\alpha=1.1$ and the downwards pointing triangles 
show the results for $\alpha=1.7$. 
The errorbars, plotted for the $\alpha=1.1$ case only, are obtained by 
taking the {\it rms} scatter from the mock catalogues and rescaling 
to the amplitude of the mean spectrum recovered from the APM data 
(see the next Section for a discussion of how the results obtained 
here differ from those of GB98, which were used to construct the 
mock catalogues). These errors are further rescaled by a factor $1/\sqrt{3}$, 
since they correspond to the error on the mean from four measurements.
The results for the power spectrum are tabulated in Table \ref{table_pow}.
The final column gives the fractional {\it rms} obtained from the mocks; 
to derive the absolute error on the measured power spectrum, the amplitude 
of the power spectrum should be multiplied by this fractional error. 
The convergence of the algorithm is illustrated in the inset in the upper 
panel of Fig. \ref{fig:apm_alpha}, which 
shows the value of the ``goodness'' function, $G$, (defined 
by eqn. \ref{eq:goodness}), for different iterations. 
The scale along the x-axis gives the iteration number relative to the 
``best'' iteration. (Note the best iteration can be different for each 
zone.) 
Negative values correspond to iterations that precede the one identified as 
the best.
There is a clear minimum in the mean ``goodness'' at the best iteration.

The lower panel of Fig. \ref{fig:apm_alpha} shows the ratio between 
the inversion results obtained for the two different values of 
$\alpha$ (open triangles).  The dotted lines show the fractional 
error on  the estimate of the APM power spectrum.  There is 
little difference between the power spectra recovered for the 
different values of $\alpha$; the power spectrum obtained for 
$\alpha=1.7$ has an amplitude that is around $10$\% higher than 
is the case when a value of $\alpha=1.1$ is adopted. 
This difference is in the sense expected because the amplitude of $P(k,z)$ 
drops more rapidly with increasing redshift in the $\alpha=1.7$ case; 
the amplitude of the angular clustering is fixed, so $P(k,z=0)$ must be 
larger for larger values of $\alpha$ to compensate for the stronger 
decline in amplitude of clustering with redshift. 
The shape of the recovered power spectrum is remarkably insensitive to 
the precise choice of $\alpha$.

\subsection{Comparison with previous estimates of the  galaxy power spectrum}
\label{sec:apmcomp}

\begin{figure}
{\epsfxsize=8.truecm 
\epsfbox[47 140 560 680]{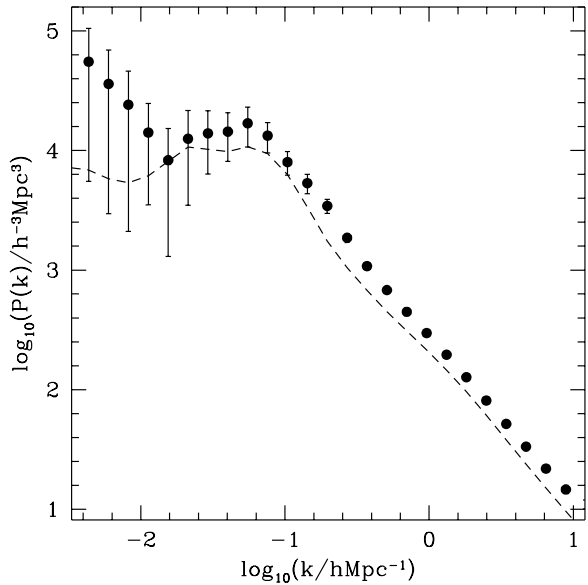}}
\caption{
The APM galaxy power spectrum averaged over the $4$ zones of 
the APM catalogue (points with error bars), and the 
estimate of GB98 (dashed line). The errorbars on our estimate
are obtained from the mock catalogues.
}
\label{fig:apm_comparison}
\end{figure}

\begin{figure}
{\epsfxsize=8.truecm 
\epsfbox[35 170 553 679]{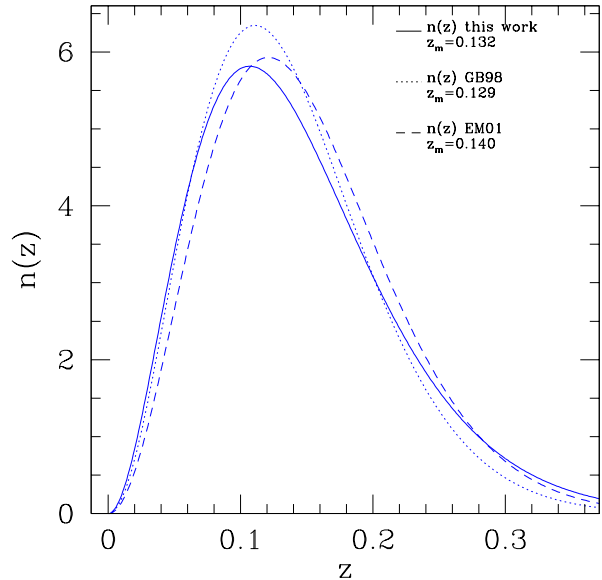}}
\caption{
Normalised redshift distributions used in our 
inversion (solid line), in GB98 (dotted line),
and EM01 (dashed line). 
}
\label{fig:nz}
\end{figure}

In this section we compare our measurement of the APM
power spectrum with other estimates of the power spectrum 
of galaxy clustering obtained from several surveys using 
a variety of methods.
We restrict our attention here to the results 
obtained using an evolution factor $\alpha=1.7$, since 
the comparisons extend to scales that are in the 
non-linear clustering regime. The errors quoted on our 
results have been divided by $\sqrt{3}$, to account for 
the fact that our estimate of the APM power spectrum is 
obtained by averaging over $4$ APM zones. 

Figure \ref{fig:apm_comparison} shows the APM power spectrum 
recovered using $\alpha=1.7$ using the algorithm described in 
Section 2 (filled points). 
The dashed line shows the $P(k)$ measurement of GB98 
for the same survey data. 
It is clear that the shape of the power spectrum 
recovered by our revised algorithm is very similar to that 
obtained by GB98. However the inflection around $\log(k/h$\rm Mpc$^{-1}) 
\sim -1$ found by GB98 is somewhat less pronounced 
in our estimate of the power spectrum. 
The inversion algorithm used by GB98 is quite similar to the one employed 
in this paper, so the origin of any discrepancy between the two estimates 
of the power spectrum lies elsewhere. 
The offset in amplitude of a factor of approximately 1.25 is 
due to the different choices made for the value of the parameter 
$\alpha$; GB assumed $\alpha=0$, whereas our results, as plotted 
in Fig. \ref{fig:apm_comparison}, are for $\alpha=1.7$.

In addition to the assumed rate of clustering evolution, a further 
possible source of discrepancy with previous work on 
the clustering of APM galaxies is in the form adopted for 
the redshift distribution of galaxies. 
We compare the new model for the redshift distribution of APM Survey 
galaxies derived in Section \ref{ssec:nofz} with previous estimates in 
Fig.  \ref{fig:nz}.
The solid line shows the fit given by eqn. (\ref{eq:nzfit}). The dotted 
line shows the fit to the redshift distribution used by GB98, which 
has a slightly less extended tail at higher redshifts, which also helps
to explain the differences between the GB98 $P(k)$ measurement and the 
one presented in the current paper. 
The dashed line shows a fit derived by Efstathiou \& Moody (2001), 
using the 11,000 galaxies taken from  high completeness regions of 
the southern 2dFGRS data at an early stage in the survey. 

\begin{figure*}
{\epsfxsize=17.truecm 
\epsfbox[60 195 580 710]{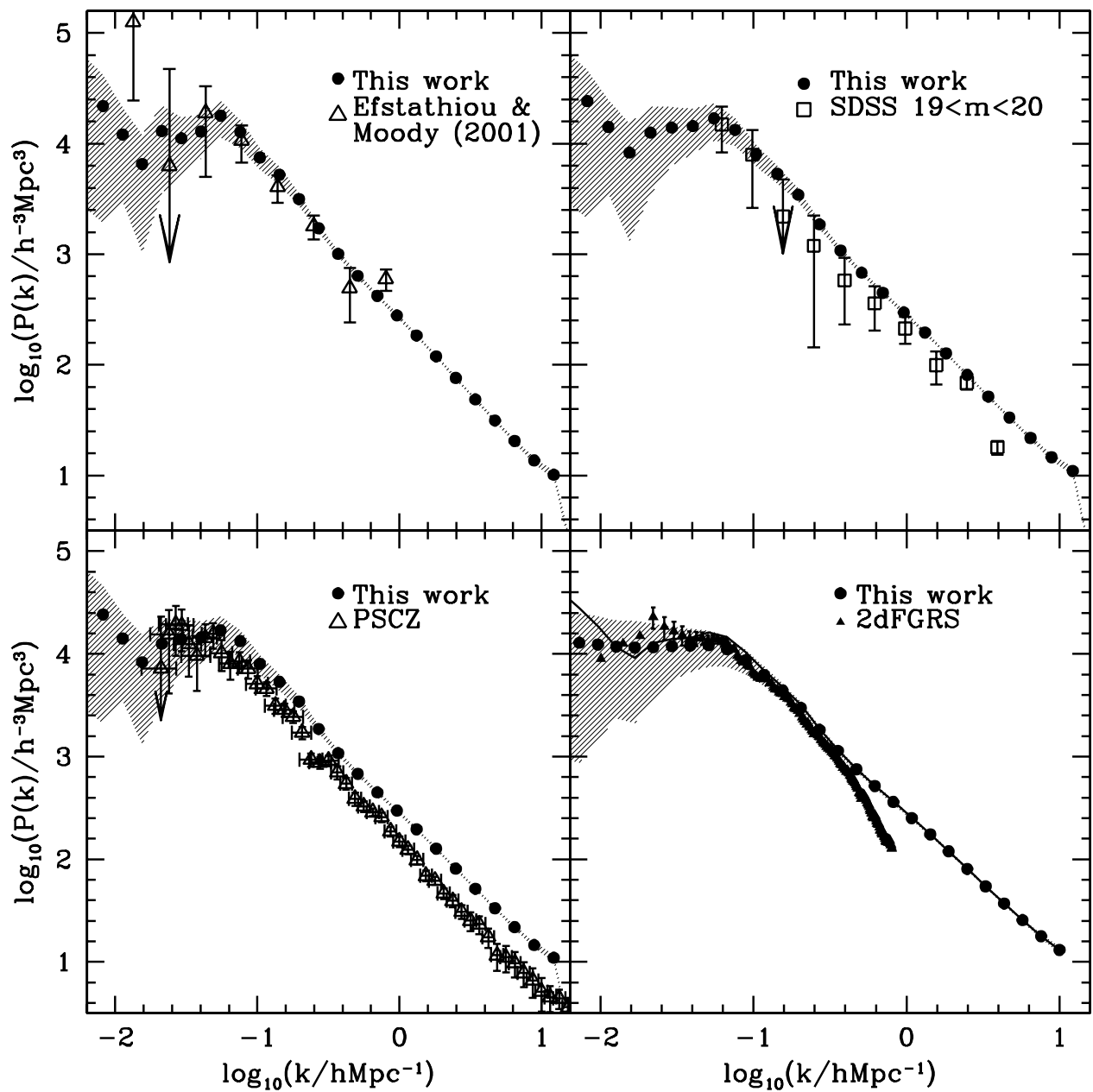}}
\caption{
Comparisons of our estimate of the APM power spectrum (filled
circles) using $\alpha=1.7$ with other estimates. 
The shaded areas show the $68\%$ confidence level
assigned to our estimate of the APM power spectrum, obtained using
the Hubble Volume APM mock catalogues.
The upper left panel shows the results
obtained by Efstathiou \& Moody (2001, open triangles) for the APM catalogue.  
The open squares  in the upper right panel
show the results obtained by Dodelson et al. (2002)
for an apparent magnitude slice $19<r^{*}<20$ extracted from early SDSS data.  
The lower left panel shows the
comparison between our estimate of the APM power spectrum
and the estimate obtained for the PSCz survey by Hamilton \& Tegmark
(2002, triangles with errorbars).
The lower right panel shows
comparison between the convolution of the 2dF window function
with our estimate of the APM power spectrum (filled circles,
the shaded area shows the $68\%$ confidence level scaled by the 
difference in amplitude induced by the convolution)
and  the estimate obtained for the 2dFGRS by Percival et al. (2001) also
affected by this convolution (triangles
with errorbars).  The solid line shows the unconvolved APM power 
spectrum. 
}
\label{fig:apmcomp}
\end{figure*}

We compare our measurement of the power spectrum of APM Survey galaxies 
with other estimates of the galaxy power spectrum in Fig. \ref{fig:apmcomp}.
In each panel, the filled circles show our measurement of the power 
spectrum of APM galaxies and the shaded region shows the error on the 
mean. In the lower right-hand panel, the APM galaxy power spectrum has 
been convolved with the window function of the 2dFGRS. 

Efstathiou \& Moody (2001) applied a maximum likelihood estimator to the 
projected counts in cells of APM galaxies to obtain the 3-dimensional 
power spectrum.
These authors assumed a Gaussian form for the likelihood, which limits the 
range of wavenumbers over which their inversion technique can be applied. 
Efstathiou \& Moody's results are shown by the open triangles in the top-left 
panel. For wavenumbers where their results can be compared with ours, 
$k<1h$ Mpc$^{-1}$, the two estimates are in excellent agreement.

The power spectrum of APM Survey galaxies is compared with 
$P(k)$ derived from an inversion of angular clustering in the 
SDSS in the top-right panel of Fig. \ref{fig:apmcomp}
(Dodelson \etal 2002).
The SDSS data used are the Early Data Release of the photometric 
catalogue, consisting of 1.5 million galaxies. (Note, as with the 2dFGRS, 
the photometric SDSS catalogue is deeper than spectroscopic sample.) 
The first step in the inversion process is the measurement of the 
angular correlation function (carried out by Connolly \etal 2002).
The power spectrum is obtained by inverting Limber's equation in 
matrix form using Singular Value Decomposition, following the scheme 
devised by Eisenstein \& Zaldarriaga (2001). 
An advantage of the approach of Dodelson \etal compared with 
that followed by Eisenstein \& Zaldarriaga is the use of mock SDSS 
surveys to estimate the errors on the measured angular correlation 
function. These mock catalogues are constructed using the PTHalos code 
described by Scoccimarro \& Sheth (2002).
The Dodelson \etal estimate of $P(k)$ has a slightly lower amplitude 
(by a factor of $1.6$) than our estimate. 
There are two main explanations of this offset: 
(i) The SDSS galaxies are selected in the $r^{*}$-band, which has 
a longer effective wavelength that the $b_{\rm J}$-band ($6216\AA$ for $r^{*}$ 
versus $4482\AA$ for $b_{\rm J}$). (ii) Different assumptions are made for 
the evolution of clustering, and the samples have different median redshifts. 
Dodelson \etal assume $\alpha=0$ in our notation. 

The lower left hand panel of Fig. \ref{fig:apmcomp} contrasts our measurement 
of the power spectrum of galaxies selected in the optical $b_{\rm J}$-band 
with an estimate of the real space power spectrum of galaxies selected 
by their 60 micron emission, as computed from the IRAS PSCz 
by Hamilton \& Tegmark (2002).
Their estimate is a composite based on two different techniques, 
one operating on linear scales and the other on scales for which the 
density fluctuations are nonlinear. 
The technique applied to linear scales assumes Gaussian fluctuations, 
and the scheme used on non-linear scales makes a plane-parallel 
approximation, with the real space power spectrum inferred from 
the redshift space power spectrum in the transverse direction.  
As remarked upon previously by many authors 
(e.g. Peacock 1997, Hoyle \etal 1999, Hamilton \& Tegmark 2002), 
there is evidence for a scale dependent {\it relative} bias 
between the power spectrum of galaxies selected by their emission 
in the optical and that of galaxies selected in the far infra-red. 
The relative bias, defined as $b = \sqrt{P_{\rm APM}/P_{\rm PSCz}}$, 
ranges from $b=1.25$ on scales around 
$k \sim 0.1$ h Mpc$^{-1}$ to $b=1.6$ at $k \sim 10$ h Mpc$^{-1}$.  
This scale dependence of the relative clustering amplitude 
indicates that APM galaxies trace the underlying mass distribution 
in a different way than PSCz galaxies do. In particular, APM galaxies 
display more clustering on small scales, which suggests that they 
occur in clusters more often than PSCz galaxies do. 
Such a difference in the clustering signals of different types of 
galaxy is a powerful constraint on models of galaxy formation. 
In terms of the halo occupation model deployed by Benson \etal (2000) 
to explain galaxy clustering, one would expect the number of pairs of 
galaxies in the optical to increase more rapidly with halo mass than is the 
case for galaxies in the far infra-red.

Finally, we compare the real space power spectrum of APM Survey 
galaxies with a direct estimate of $P(k)$ in redshift space  
made from the 2dFGRS (Percival \etal 2001) in the lower right-hand 
panel of Fig. \ref{fig:apmcomp}.
Percival \etal used 160000 galaxies to measure $P(k)$ using a 
fast Fourier transform. Their result is influenced by the 
finite width of the survey window function in Fourier space.
To permit a comparison on a more equal footing, we have convolved the 
APM galaxy power spectrum with the window function of the 2dFGRS. 
The result after convolution is shown by the filled points in this panel; 
the original, unconvolved spectrum is shown by the solid line for reference. 
We note that the errorbars in the 2dFGRS estimate are much smaller 
than our $68\%$ confidence level range; however, it should be 
remembered that the 2dFGRS $P(k)$ points are correlated, and so the 
errorbars plotted here, showing the diagonal component of the 
covariance matrix, do not give the full picture.
The two measurements of $P(k)$ are in good agreement. There are several 
factors that could lead to differences between the real space and 
redshift space power spectra: (i) The distortion of the clustering 
pattern arising from the gravitationally induced peculiar motions of 
galaxies. Peculiar motions result in an enhancement of clustering 
on large scales and a damping of power on small scales 
(e.g. see Fig. 4 of Padilla \& Baugh 2002). Peculiar motions are responsible 
for the steepening of the redshift space power spectrum at $\log(k/h{\rm Mpc}^{-1})> -0.5$ in Fig. \ref{fig:apmcomp}. 
(ii) The radial weighting scheme employed by Percival \etal to obtain a
minimum variance estimate of $P(k)$ means that the characteristic luminosity 
of galaxies in the 2dFGRS sample is $\sim 2L_{*}$ compared with $L_{*}$ in the 
APM Survey. Norberg \etal (2001, 2002b) found that 2dFGRS galaxies display a 
linear dependence of clustering strength on luminosity. 
(iii) The clustering signal measured by Percival \etal corresponds to 
$P(k)$ at the median redshift of the sample; our $P(k)$ result is corrected 
to the present day, by a factor that depends upon the parameter $\alpha$ 
(see eqn. \ref{eq:alpha}).

\section{Constraints on models of structure formation} 
\label{sec:constraints}

In this section we examine the constraints upon theoretical models 
of structure formation that result from our measurement of 
the real space galaxy power spectrum. 
We begin by making a general comparison between our results and what 
has become known as the ``concordance'' cold dark matter model (Ostriker 
\& Steinhardt 1995), before 
going on to present formal constraints on the parameters of the CDM 
model. 

\subsection{Comparison with the Cold Dark Matter concordance model}
\label{ssec:comp_cmb}
\begin{figure}
{\epsfxsize=8.truecm 
\epsfbox[50 180 400 680]{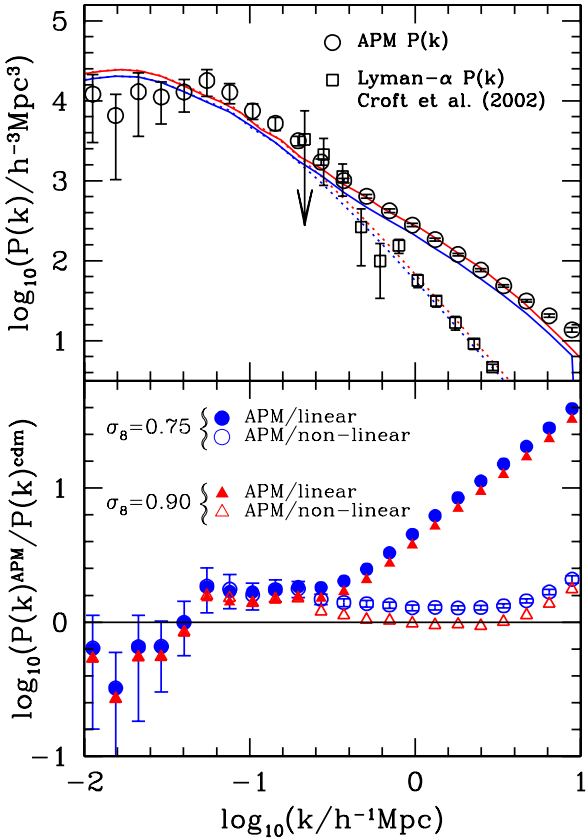}}
\caption{
The upper panel shows the APM power spectrum (open circles with errorbars). 
The squares show an estimate of the linear mass power spectrum at $z=2.5$, 
rescaled to the present day, assuming $\Omega_{m}=0.3$ and $\Omega_{\Lambda}=0.7$, 
taken from Croft \etal (2002).
The dotted lines show the best fit CDM models to the spectrum of 
temperature fluctuations in the microwave background radiation as deduced by  
Netterfield \etal (2002); the lower amplitude dotted line is for 
$\sigma_{8}=0.75$ and the other is for $\sigma_{8}=0.90$. 
The solid lines show the corresponding  nonlinear mass power spectra. 
In the lower panel, the ratio of the APM power spectrum to the various 
theoretical models is plotted; the filled symbols show the ratio to the 
linear theory $\Lambda$CDM power spectra and the open symbols show the 
ratios to the nonlinear power spectra. 
}
\label{fig:apm_ly}
\end{figure}

The $\Lambda$CDM model, with around $30\%$ of the critical density of 
the Universe made up by cold dark matter and baryons, but with a 
spatially flat geometry due to the presence of a non-zero cosmological 
constant, has enjoyed a number of successes. Originally motivated to 
explain the excess clustering seen at large angular separations in the 
APM survey over the predictions of the ``standard'' CDM model of the 1980's  
(Efstathiou, Sutherland \& Maddox 1990), the $\Lambda$CDM model is also 
consistent with the location of the Doppler peaks in the power spectrum 
of microwave background temperature fluctuations (de Bernardis \etal 2000), 
and with the Hubble diagram of high redshift supernovae 
(Perlmutter \etal 1999). 

The detection of features in the power spectrum of microwave background 
fluctuations due to acoustic oscillations has allowed the parameters 
of the $\Lambda$CDM concordance model to be firmed up 
(e.g. Jaffe \etal 2001; Percival \etal 2002).
In Figure \ref{fig:apm_ly}, we compare the power spectrum of APM galaxies 
with a linear theory power spectrum that is the best fit to the Boomerang 
experiment data for prior assumptions that the Universe is flat 
and that the values of the density parameter and cosmological constant 
are consistent with the observations of high redshift supernovae 
(taken from Table 5 of Netterfield \etal 2002).
Two versions of the linear theory spectrum are shown (dotted lines); 
one corresponding to a normalisation of $\sigma_{8}=0.75$, which is 
the best fit to the CMB data (Efstathiou \etal 2002) and the other 
to $\sigma_{8}=0.93$, as suggested by the abundance of hot X-ray 
clusters (e.g. Eke, Cole \& Frenk, 1996).
The linear theory spectra were computed using CMBFAST (Seljak \& Zaldarriaga 
1996).
The theoretical predictions are in reasonably good agreement with the 
shape of the APM real space power spectrum down to scales around 
$\log(k/h{\rm Mpc}^{-1}) \sim -0.5$.
It is instructive to also plot the power spectrum inferred from the 
Lyman-alpha forest by Croft \etal (2002) in this comparison (see also 
Fig. 19 of Croft \etal 1999). Croft \etal argue that their technique recovers 
the linear theory power spectrum of mass fluctuations at high 
redshift ($z \sim 3$); there is, however, some debate about the interpretation 
of the power spectrum results (Gnedin \& Hamilton 2002; 
Zaldarriaga, Scoccimarro \& Hui 2001). Despite these reservations, it 
is suggestive that the most recent 
Croft \etal results, when extrapolated to the present 
day using the cosmological parameters of the 
concordance model,  lie on top of the linear theory $\Lambda$CDM predictions, 
forming a smooth transition to those APM $P(k)$ points that are still in the 
linear regime.

On smaller scales (or higher wavenumbers), the power spectrum of APM 
galaxies disagrees with the linear theory predictions. This is largely 
the result of nonlinear evolution of density fluctuations, which is 
illustrated by the solid lines in Fig. \ref{fig:apm_ly}, 
which show the nonlinear power spectrum computed using the 
transformation described by Smith \etal (2002). The remaining differences 
between the nonlinear power spectrum of the concordance model and the 
galaxy power spectrum could be interpreted as a scale dependent bias 
factor. This point is emphasised in the lower panel of Fig. \ref{fig:apm_ly}, 
in which we plot the ratio of the APM galaxy power spectrum to various  
theoretical power spectra. The filled symbols show the ratio of the 
APM galaxy power spectrum to the linear theory $\Lambda$CDM $P(k)$; the 
circles show the ratio for $\sigma_{8}=0.75$ and the triangles show the 
case for which $\sigma_{8}=0.9$. The open symbols show the ratio of 
the galaxy power spectrum to the nonlinear $\Lambda$CDM models. 
At high wavenumbers, accounting for nonlinear evolution dramatically reduces 
the discrepancy between the theoretical predictions and the measured 
galaxy power spectrum.

\subsection{Constraints on CDM model parameters} 
\label{ssec:parameters}



\begin{figure}
{\epsfxsize=8.truecm 
\epsfbox[47 27 523
520]{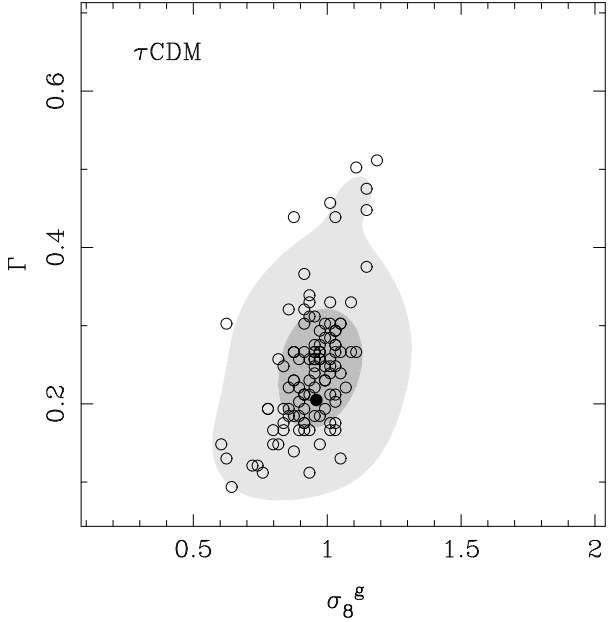}}
\caption{
The distribution of best fit parameters in the ($\Gamma, \sigma^{g}_{8}$) 
plane for mock APM zones taken from the $\tau$CDM Hubble Volume simulation. 
Each point corresponds to the best fit parameters for the power 
spectrum recovered from one zone. The darker shaded region encloses 
the results from $68\%$ of the mocks, the light shading 
extends this region to enclose $95\%$ of the mocks.
The filled circle shows the best fit 
values of $\Gamma$ and $\sigma^{g}_{8}$ for the galaxy power spectrum 
measured by FFT in the full simulation box. 
}
\label{fig:l}
\end{figure}

\begin{figure}
{\epsfxsize=8.truecm 
\epsfbox[47 26 523
520]{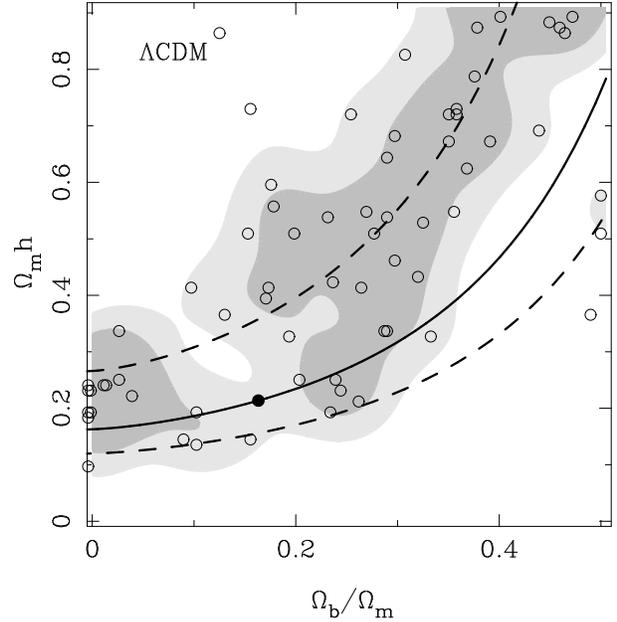}}
\caption{
The distribution of best fit parameters in the 
($\Omega_{b}/\Omega_{m}$, $\Omega_{m} h$) plane for mock APM zones, 
this time taken from the $\Lambda$CDM Hubble Volume simulation. 
Each point corresponds to the best fit parameters for the power 
spectrum recovered from one mock zone. The heavy shaded 
region encloses the results from $68\%$ of the mocks, the light shading 
extends this region to enclose $95\%$ of the mocks.
The solid line shows the locus expected in this plane as the baryon fraction 
is increased for a fixed power spectrum shape, as explained in the text; 
the dashed lines show the effects of the uncertainty in $\Gamma$.
The filled circle shows the best fit 
values of $\Omega_{b}/\Omega_{m}$ and 
$\Omega_{m}$ for the galaxy power spectrum 
measured by FFT in the full simulation box. 
}
\label{fig:l2}
\end{figure}

\begin{figure}
{\epsfxsize=8.truecm 
\epsfbox[47 26 523 517]{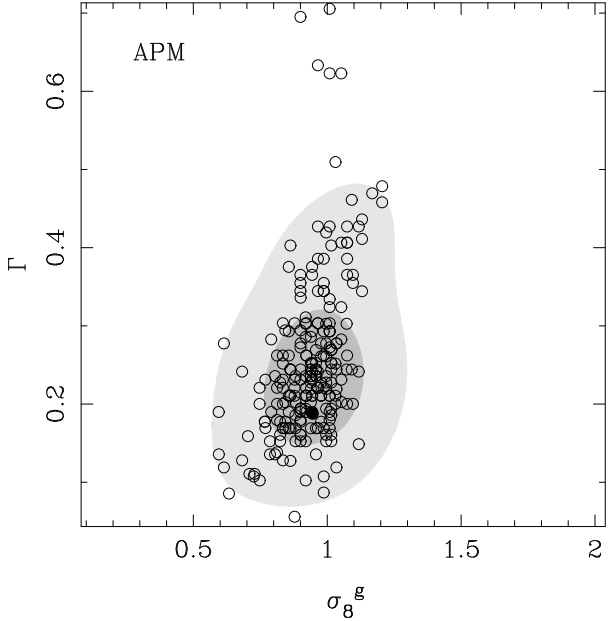}}
\caption{
The constraints in the ($\Gamma, \sigma^{g}_{8}$) parameter space 
from the power spectrum of APM Survey galaxies.
The contours show the $68\%$ (heavy) and $95\%$ (heavy plus light) 
confidence intervals, estimated using mock APM catalogues.
The filled circle shows the best fit 
values of $\Gamma$ and $\sigma^{g}_{8}$ for the galaxy power spectrum 
measured by FFT in the full simulation box. 
}
\label{fig:l_apm}
\end{figure}

\begin{figure}
{\epsfxsize=8.truecm 
\epsfbox[47 26 523 517]{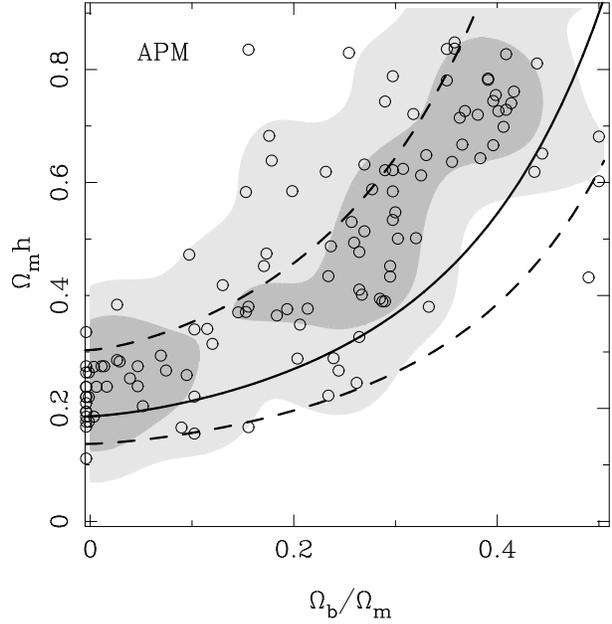}}
\caption{
The constraints in the ($\Omega_{b}/\Omega_{m}$, $\Omega_{m} h$) 
parameter space 
from the power spectrum of APM Survey galaxies.
The contours show the $68\%$ (heavy) and $95\%$ (heavy plus light) 
confidence intervals, estimated using mock APM catalogues.
}
\label{fig:l2_apm}
\end{figure}

We now generalise the comparison presented in the previous section to 
examine the constraints placed upon the parameters of the CDM model 
of structure formation by the measurement of the galaxy power spectrum. 

In order to simplify the interpretation of the galaxy power spectrum, 
we restrict our attention to scales on which the underlying density 
fluctuations are expected to be in the linear regime. We therefore 
only consider the power spectrum at wavenumbers $k<0.25h $Mpc$^{-1}$. 
The variance in fluctuations in the galaxy distribution when smoothed on 
this scale is $\sigma = 0.1$; Baugh \& Efstathiou (1994b) demonstrated 
that linear perturbation theory still gives an accurate description 
of the shape of the power spectrum under this condition.
A further consequence of focusing on the largest scales is that the 
relation between fluctuations in the distribution of galaxies and 
of the dark matter is expected to be a simple scaling in amplitude 
(e.g. Cole \etal 1998).
In this Section, we use the power spectrum of APM Survey galaxies 
obtained for an evolution of clustering described by $\alpha=1.1$, 
which we argued is appropriate for the linear regime in the concordance 
$\Lambda$CDM model (see Section \ref{sec:app}).

We consider two sets of parameters that can be used to describe the 
shape of the power spectrum: 
\begin{itemize}
\item[(i)] 
The amplitude of fluctuations in spheres of radius $8 h^{-1}$Mpc,
$\sigma^{g}_{8}$, and the power spectrum shape parameter, $\Gamma$, 
as defined by Efstathiou, Bond \& White (1992).
Note that we assume $\sigma^{g}_{8} = b 
\sigma_{8}$, where $b$ is a scale independent bias factor (which 
applies for $k<0.25h $Mpc$^{-1}$).  
\item [(ii)]
The present day mass fraction of baryons, $\Omega_{b}/\Omega_{m}$, 
and the density parameter times the Hubble constant in units of 
$100 \rm{km}/\rm{s}/\rm{Mpc}$, $\Omega_{m} h$. 
\end{itemize}
Theoretical power spectra are computed for a set of grid points in these 
parameter spaces using the physically motivated fitting formula given 
by Eisenstein \& Hu (1998); we have checked that this formulation 
agrees with the output of CMBFAST to better than $10\%$ over the 
range of wavenumbers considered.

The best pair of parameters that describe a measured power spectrum, 
$P^{m}$, is found by minimising $\chi^2$;
\begin{equation}
\chi^2=\sum_i (P^{m}(k_i)-P^{CDM}(k_i))^2/\sigma_i^2
\label{eq:chi}
\end{equation} 
where the index $i$ runs over the wavenumber bins $k_i$ up to 
$k=0.25h $Mpc$^{-1}$, and $P^{CDM}$ is the theoretical CDM power spectrum. 
The uncertainty in the measured power spectrum, 
$\sigma_i$, is the {\it rms} scatter obtained from the 
Hubble Volume mocks (see Section \ref{ssec:error_estimation}).

The outcome of this procedure is a pair of CDM parameters that best
matches the measured $P(k)$. In order to find the variance on the 
estimate of the best fit parameters, we repeat this process 
for the power spectra recovered from each mock APM zone.
The result is a set of points in the two parameter spaces, either  
($\Gamma, \sigma^{g}_{8}$) or ($\Omega_{b}/\Omega_{m}$, $\Omega_{m} h$).
The distribution of points is smoothed with a Gaussian filter of width 
comparable to the spacing of the grid points in each parameter. 
The smoothed distribution is used to define contours. These contours are 
then used to delineate 
regions within which the number of points can be counted 
to give the $68\%$ and $95\%$ confidence intervals.
This process is illustrated in \ref{fig:l}, in which we show the 
constraints in the ($\Gamma, \sigma^{g}_{8}$) plane for the $\tau$CDM mock 
catalogues. Each point corresponds to the best fit pair of parameters 
for the power spectrum recovered from a mock APM zone. The dark shading 
encloses 68\% of the points, and the light shading extends this to 
include 95\% of the points. 
The filled circle shows the best fit 
values of $\Gamma$ and $\sigma^{g}_{8}$ for the galaxy power spectrum 
measured by FFT in the full simulation box. 
The constraints are less 
strong in the ($\Omega_{b}/\Omega_{m}$, $\Omega_{m} h$) plane, which is 
plotted in Fig. \ref{fig:l2_apm}. In this example, we show the 
results for the $\Lambda$CDM simulation mocks since the initial conditions 
for the $\Lambda$CDM Hubble Volume simulation were generated using 
CMBFAST with a baryon fraction of $\Omega_{b}/\Omega_{m}=0.16$, 
as indicated by the filled circle.
The solid curve defines the locus expected in the parameter space to 
produce a fixed power spectrum shape (for $k<0.25h $Mpc$^{-1}$) as the 
baryon fraction is increased; the dashed lines show how the locus 
of the parameter space covered in practise due to the uncertainty in 
$\Gamma$. 
This curve is a fit to $\Lambda$CDM power spectra generated 
using the Eisenstein \& Hu (1998) formula for the CDM transfer 
function for an $\Omega_{m}=0.3$ and $\Omega_{\Lambda}=0.7$ cosmology, 
and can be written as: 
{\small
$$
\Omega_{m} h= \Gamma (1^{+0.6}_{-0.2}) \exp \left[
5.9 (1 + 1.34^{+0.80}_{-0.33} \Gamma)  
\left( \frac{\Omega_b}{\Omega_{m}}\right)^{1.65+1.23\frac{\Omega_b}{\Omega_{m}}} \right].
$$}

We now apply these parameter constraints to the measured APM 
Survey power spectrum, using the mock catalogues to define the 
errors on the parameters. 
As the mock catalogues were constructed to reproduce an earlier 
measurement of the power spectrum of APM galaxies, a small rescaling 
is required in the parameter space, to shift the error contours to 
lie around the best fitting parameters for our measurement of 
the power spectrum.
The results in the ($\Gamma, \sigma^{g}_{8}$) plane are shown in 
Fig. \ref{fig:l_apm}, with best fit values obtained of 
$\sigma^{g}_{8}=0.96^{+0.17}_{-0.20}$ and 
$\Gamma= 0.19^{+0.13}_{-0.04}$. 
Efstathiou \& Moody (2001) find similar constraints on these 
parameters: these authors quote the $2-\sigma$ ranges 
of $0.78 \leq (\sigma^{g}_8) \leq 1.18$, and $0.05 \leq \Gamma \leq 0.38$.  
Fig. \ref{fig:l2} shows the constraints derived from the APM galaxy 
power spectrum in the ($\Omega_{b}/\Omega_{m}$, $\Omega_{m} h$) plane.
Percival \etal (2001) performed a similar analysis using the redshift 
space power spectrum of 2dFGRS. The shape of the contours in the 
($\Omega_{b}/\Omega_{m}$, $\Omega_{m} h$) plane found by Percival \etal 
are similar to ours, with slightly more curvature for larger values of 
$\Omega_{m}h$. Percival \etal find somewhat narrower $68\%$ and $95\%$ 
contours than we do. This is to be expected as a direct estimate of $P(k)$ 
in three dimensions samples more Fourier modes than is the case for 
projected data, though the larger volume covered by the APM Survey 
compared with the 2dFGRS offsets this to some extent.

\section{The real space correlation function of APM galaxies}
\label{sec:xi}

\begin{figure}
{\epsfxsize=8.truecm 
\epsfbox[20 170 570 694]{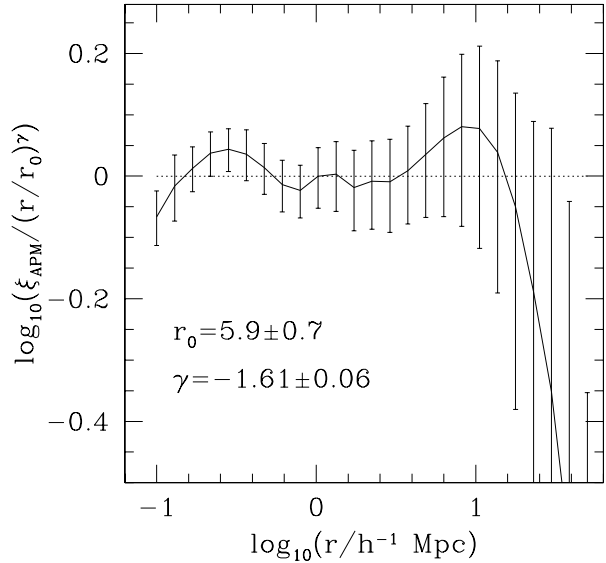}}
\caption{
The real space correlation function estimated for APM Survey 
galaxies using $\alpha=1.7$,
divided by a reference power law, with the parameters 
shown in the legend.
}
\label{fig:xishape}
\end{figure}

\begin{figure*}
{\epsfxsize=16.truecm \epsfysize=14.truecm 
\epsfbox[40 200 583 680]{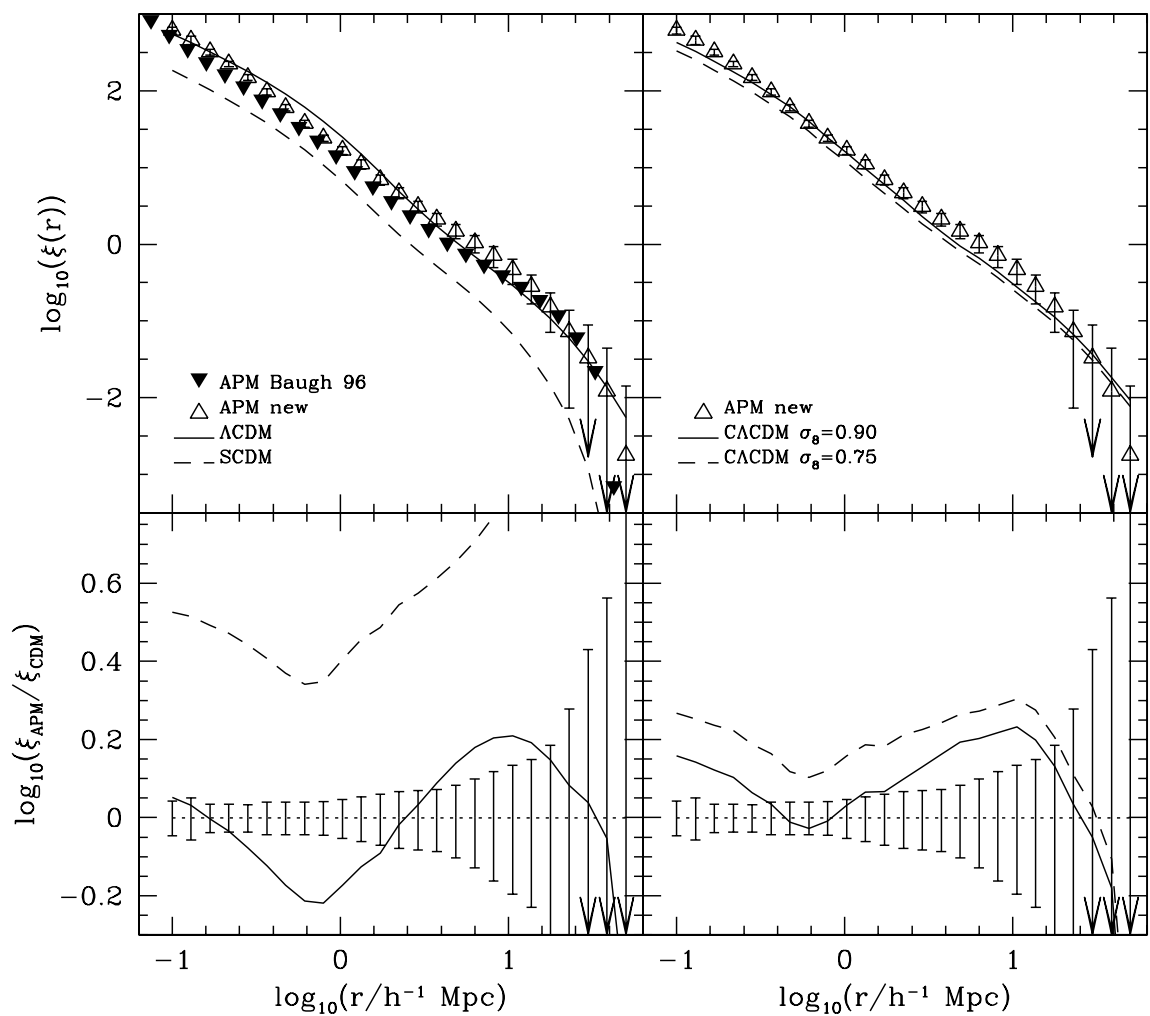}}
\caption{The real space correlation function of APM Survey 
galaxies.
The upper left hand panel compares our estimate of the 
correlation function (open triangles) with an earlier estimate from 
Baugh (1996 - filled triangles).  
The errorbars show the {\it rms} scatter from the mocks.
We also show, in the same panel, the correlation functions for 
a SCDM (dashed line) and a $\Lambda$CDM (solid line) models. 
The lower left panel shows ratios between the APM correlation function
estimated in this paper to the SCDM and $\Lambda$CDM
correlation functions, using the same line types.
The errorbars indicate the amplitude of relative errors as 
a function of scale $r$.
The right hand side panels also show the present estimate of the
APM correlation function, but in this case it is compared to
the best CDM fits to Boomerang results (Netterfield et al. 2002),
for $\sigma_8=0.75$ (dashed line) and $\sigma_8=0.9$ (solid line). 
Using the same line styles, the ratios
between the APM and CDM correlation functions are shown in the
lower right panel.
}
\label{fig:xi}
\end{figure*}

The two-point galaxy function is still a popular way of characterising 
galaxy clustering, particularly on small scales. Though it is somewhat 
less appealing than the power spectrum from a theoretical point of view, 
the correlation function has the advantage of being simpler conceptually. 
As the power spectrum and correlation function are Fourier transforms of 
one another, we can use the measurement of the APM galaxy power spectrum 
to infer the galaxy correlation function in real space. 
An independent approach is to write Limber's equation explicitly 
in terms of the correlation function, and invert this using 
the techniques outlined in Section 2 (Baugh 1996).
Another possibility is to numerically invert the projected spatial 
correlation function (Saunders, Rowan-Robinson \& Lawrence 1992; 
Hawkins \etal 2002).

For an isotropic power spectrum, the correlation function is given by:  
\begin{equation}
\xi(r)=\frac{1}{2\pi^{2}}\int_0^{\infty}P(k)\frac{ \sin(kr)}{kr} k^2 {\rm d}k,
\label{eq:xi}
\end{equation}
where $r$ is the comoving spatial separation.
To avoid spurious numerical artifacts (ringing), 
the measured power spectrum is extrapolated to 
high wavenumbers by assuming a suitable power law.
We estimate the correlation function from the measured APM power 
spectrum obtained for both clustering evolution scenarios, 
$\alpha=1.1$ and $\alpha=1.7$, and give the results in 
table \ref{table_xi}.  The relative errors in this
table are the {\it rms} scatter over the correlation functions 
obtained from the Fourier transform of the power spectra of
the $260$ mock APM zones.

The real space two point correlation function of APM Survey galaxies
is well described by a power law, $\xi(r) = (r/r_{0})^{\gamma}$, with 
$r_0=5.9 \pm 0.7$ and $\gamma =-1.61 \pm 0.06$, as 
shown in Fig. \ref{fig:xishape} (here results are shown for 
the $\alpha=1.7$ case). There is a suggestion that 
the correlation function rises above a power law for pair separations 
around $r \sim 10 h^{-1}$Mpc and then falls below the power law, 
as expected in CDM models. However, the size of the errors on our 
measurement of the correlation function prevents us from making   
firm statements about these features with any conviction. 
The correlation function we find is in very good agreement with 
an independent estimate from the 2dFGRS by Hawkins \etal (2002 - see 
their Fig. 11).
These authors use a different inversion technique to estimate the 
real space correlation function from the projected correlation function 
of the 2dFGRS. Hawkins \etal fit a power-law to their estimate of 
the real space correlation function and obtain a correlation length 
of $r_{0} = 5.05 \pm 0.26 h^{-1}$Mpc and a slope of $\gamma = -1.67 \pm 0.03$.

Figure \ref{fig:xi} shows the correlation function for the case where 
$\alpha=1.7$ (open triangles).  In the left-hand panel, we compare 
this result for $\xi(r)$ with the 
previous result for the spatial correlation function of APM Survey 
galaxies, derived by Baugh (1996: filled triangles).
The amplitude of our estimate of the correlation function is 
somewhat higher in amplitude, particularly for $1<r/h^{-1}$Mpc$<10$, as 
expected from the comparison of the measurements of the power spectrum 
shown in Fig. \ref {fig:apmcomp}.
Also in the left-hand panel, we compare the measurements of the galaxy 
correlation function with the theoretical predictions from two flavours of 
CDM model. The dashed line shows the nonlinear correlation function of the 
mass in a  standard CDM model (SCDM $\Omega_{m}=1$, $\Gamma=0.5$, $\sigma_{8}=0.5$) and the solid line shows the corresponding result for $\Lambda$CDM 
($\Omega_{m}=0.3$, $\Omega_{\Lambda}=0.7$, $\Gamma=0.21$, $\sigma_{8}=0.9$).
The predictions of the SCDM model lie substantially below the measured 
galaxy correlation function; the ratio between the measured and theoretical 
correlation function is shown in the bottom panel. A significant, scale 
dependent bias factor would be required to reconcile the SCDM model 
with the measured correlation function. 
Furthermore, the bias would be required to increase with pair separation, 
which emphasises that this CDM model has the wrong shape of correlation 
function. 
The earlier APM results were used 
to motivate the need for an anti-bias at small separations, i.e. 
a reduction in galaxy clustering relative to the dark matter in the 
$\Lambda$CDM model (Gazta\~{n}aga 1995, Jenkins \etal 1998). 
The comparison with the new correlation function estimate shows that the 
anti-bias required is much less dramatic than before.

The bottom line on bias changes again if we use a more realistic model 
for the power spectrum than the simple $\Gamma$ model. The right-hand panel 
of Fig. \ref{fig:xi} compares our measurement of the correlation function 
with the theoretical predictions for the concordance model used in 
Section \ref{ssec:comp_cmb} (to recap, $\Omega_{m}=0.27$,
$\Omega_{\Lambda}=0.67$, $\Omega_{b}=0.05$). These predictions are 
generated by first producing a linear theory power spectrum with CMBFAST, 
computing the corresponding nonlinear spectrum using the procedure 
described by Smith \etal (2002) and finally Fourier transforming the 
nonlinear mass power spectrum.    
A normalisation for the amplitude of density fluctuations needs 
to be specified; we show results for two cases, $\sigma_{8}=0.75$ 
(dashed line) and $\sigma_{8}=0.9$ (solid line). The lower panel 
shows that there is no evidence for any anti-bias relative to these 
models.  There is a slight dip in the ratio between the
APM and C$\Lambda$CDM correlation functions at a pair separation just 
below $1h^{-1}$Mpc, which is similar to the feature seen in the correlation 
function of SDSS galaxies (on a sightly larger scale), as reported 
by Zehavi \etal 2003. Zehavi \etal interpret this feature as marking 
the transition between the regime in which correlations between 
the galaxies in a single halo dominate the clustering signal, to that 
in which correlations between haloes dominate.

\begin{table}
\caption{\small{The real space correlation function of APM Survey 
galaxies. 
}}
\begin{tabular}{cccc}
\hline
\hline
\noalign{\vglue 0.2em}
$r/h^{-1}$ Mpc   & $\xi(r)$              & $\xi(r)$              & relative \\
 		 & $(\alpha=1.1)$        & $(\alpha=1.7)$        & errors\\
\noalign{\vglue 0.2em}
\hline
\noalign{\vglue 0.2em}
       0.10  &     578. &      615. &     0.10 \\
       0.13  &     428. &      455. &     0.12 \\
       0.17  &     301. &      321. &    0.084 \\
       0.22  &     210. &      223. &    0.083 \\
       0.28  &     140. &      149. &    0.080 \\
       0.36  &     90.9 &      96.6 &    0.095 \\
       0.47  &     56.9 &      60.5 &    0.095 \\
       0.61  &     35.1 &      37.4 &    0.097 \\
       0.79  &     22.6 &      24.1 &    0.099 \\
       1.03  &     15.7 &      16.8 &     0.11 \\
       1.33  &     10.4 &      11.1 &     0.13 \\
       1.73  &     6.54 &      6.97 &     0.15 \\
       2.24  &     4.40 &      4.70 &     0.16 \\
       2.90  &     2.90 &      3.09 &     0.17 \\
       3.75  &     1.99 &      2.12 &     0.18 \\
       4.86  &     1.39 &      1.49 &     0.21 \\
       6.30  &    0.976 &      1.04 &     0.26 \\
       8.16  &    0.674 &     0.716 &     0.31 \\
       10.6  &    0.447 &     0.469 &     0.36 \\
       13.7  &    0.274 &     0.282 &     0.41 \\
       17.7  &    0.150 &     0.151 &     0.53 \\
       23.0  &   0.0722 &    0.0725 &     0.90 \\
       29.8  &   0.0313 &    0.0328 &      1.7 \\
       38.6  &   0.0111 &    0.0121 &      2.6 \\
       50.0  &   0.0014 &    0.0018 &      7.0 \\
\noalign{\vglue 0.2em}
\hline
\hline
\end{tabular}\label{table_xi}
\end{table}

\section{Conclusions}
\label{sec:conc}

We have revised and extended the algorithm introduced by Baugh \& Efstathiou 
(1993) to numerically invert Limber's equation. A key ingredient of our 
approach is the use of realistic mock catalogues. These have been used 
both to test the inversion procedure (extending the work of  
Gazta\~{n}aga \& Baugh 1998) and to estimate the errors on the recovered 
power spectrum or on the value of the parameters of structure formation 
models. The Lucy inversion technique is competitive with other algorithms 
described in the literature; it is simple, fast and robust and 
can be applied over  a wide range of scales because it is not 
underpinned by the assumption of Gaussianity of density fluctuations.   

We have also addressed a number of outstanding issues that were 
raised by the earlier inversion papers.
We presented a new model for the redshift distribution of APM 
Survey galaxies, based upon the latest results from the 2dFGRS 
(Norberg \etal 2002a). We have also explored the expected evolution 
of clustering, and put forward arguments to support two particular 
choices for the value of the parameter $\alpha$ that quantifies the 
evolution. The assumption that we 
make about how the power spectrum changes with redshift is undoubtedly 
an oversimplification, but we have illustrated that the systematic errors 
incurred through this are typically at most $10\%$ for the case of the 
APM Survey. 

The main impact of the mock catalogues has been in the determination 
of the errors on our measurement of the power spectrum. The errorbars 
that we infer are comparable to those inferred by the most recent 
studies, but can also be obtained at wavenumbers for which Gaussianity 
is a poor approximation. The errors that we find on the power spectrum on 
large scales are somewhat larger than those estimated from the scatter 
between just four measurements by Baugh \& Efstathiou (1993, 1994a). 

We have used our measurement of the galaxy power spectrum 
to place constraints upon the main 
parameters that determine the power spectrum in the CDM model. 
The constraints we find on the amplitude of fluctuations in the 
galaxy density, measured in spheres of radius $8h^{-1}$Mpc, 
$\sigma^{g}_{8}=0.96^{+0.17}_{-0.20}$ and on the shape of 
the power spectrum, $\Gamma = 0.19^{+0.13}_{-0.04}$,  are 
comparable to previous results obtained for the APM Survey by 
Efstathiou \& Moody (2001).
Dodelson et al. (2002) inverted the angular clustering in an early 
data release of the SDSS and found $\Gamma=0.14^{+0.11}_{-0.06}$, 
in agreement with our constraint. 
Szalay et al. (2001) used a KL estimate of the power spectrum from 
the SDSS to derive a similar range of values: 
$\Gamma=0.188 \pm 0.04$ and $\sigma^{g}_{8}=0.915 \pm 0.06$
for a flat universe with cosmological constant.
A consensus seems to be emerging on the shape and amplitude of the 
galaxy power spectrum for bright, optically selected galaxies, 
as remarked upon by Hoyle \etal (1999).

The galaxy power spectrum that we measure is in good agreement with the 
shape of the linear perturbation theory power spectrum of the ``concordance'' 
$\Lambda$CDM model on large scales. On small scales, the form of the 
galaxy power spectrum is strongly discrepant with the linear theory 
mass spectrum. A large part of this disagreement is accounted for when 
the nonlinear evolution of the density field is taken into consideration. 
The remaining differences between the galaxy power spectrum and the 
nonlinear power spectrum of the mass are at present being 
studied with the aid of hierarchical models 
of galaxy formation. It will be of particular interest to study how the 
relationship between the distribution of galaxies and the underlying 
mass evolves with redshift by applying the technique set out in this 
paper to narrow redshift slices of galaxies selected using 
photometric redshifts from forthcoming, deep, multi-passband surveys.

\section*{Acknowledgments}
This work was supported in part by a British Council grant 
for exchanges between Durham and Cordoba, 
by CONICET, Argentina, and by a PPARC rolling grant at the University 
of Durham.  CMB was supported by a Royal Society University
Research Fellowship.  NDP acknowledges receipt of a Royal 
Society Visiting Fellowship. We would like to thank Enrique 
Gazta\~{n}aga for several helpful e-mail exchanges, Steve 
Moody for providing his results in electronic form, and
the anonymous referee for her helpful report.


\begin{thebibliography}{}
\bibitem[]{10}
Allgood, B., Blumenthal,G. \& Primack,J.R., 2001, conference 
proceeding for "Where's the Matter", Marseille.
\bibitem[]{20}
Baugh, C.M., 1996, MNRAS, 280, 267.
\bibitem[]{40}
Baugh, C.M.  \& Efstathiou, G., 1993, MNRAS, 265, 145.
\bibitem[]{50}
Baugh, C.M. \& Efstathiou, G., 1994a, MNRAS, 267, 323.
\bibitem[]{55}
Baugh, C.M. \& Efstathiou, G., 1994b, MNRAS, 270, 183.
\bibitem[]{60}
Baugh, C.M., Gazta\~naga, E. \& Efstathiou, G., 1995, MNRAS, 274, 1049.
\bibitem[]{65}
Baugh, C.M., Benson, A.J., Cole,S., Frenk, C.S., Lacey, C.G.,
1999, MNRAS, 305, L21.
\bibitem[]{70}
Benson, A.J., Cole, S., Frenk, C.S., Baugh, C.M., Lacey, C.G.,
 2000, MNRAS, 311, 793.
\bibitem[]{75}
Cole, S., Hatton, S., Weinberg, D. H. \& Frenk, C. S.,
1998, MNRAS, 300, 945.
\bibitem[]{80}
Colless, M. et al. (the 2dFGRS Team), 2001, MNRAS, 328, 1039.
\bibitem[]{90}
Connolly, A. et al. (the SDSS Collaboration), 2002, ApJ, 579, 42.
\bibitem[]{95}
Croft, R.A.C., Weinberg, D.H., Pettini, M., Hernquist, L., Katz, N.,
1999, ApJ, 520, 1.
\bibitem[]{100}
Croft, R.A.C., Weinberg, D.H., Bolte, M., Burles, S., Hernquist, L., 
Katz, N., Kirkman, D. \& Tytler, D., 2002,  ApJ, 581, 20.
\bibitem[]{105}
de Bernardis, P. et al., 2000, Nature, 404, 955.
\bibitem[]{110}
Dodelson, S. \& Gazta\~naga, E., 2000, MNRAS, 312, 774.
\bibitem[]{120}
Dodelson, S. et al. (the SDSS Collaboration), 2002, ApJ, 572, 140.
\bibitem[]{123}
Efstathiou, G., Sutherland, W. J. \& Maddox, S. J., 1990, Nature, 348, 705.
\bibitem[]{125}
Efstathiou, G., Bond, J.R. \& White, S.D.M., 1992, MNRAS, 258, 1P.
\bibitem[]{130}
Efstathiou, G. \& Moody, S.J., 2001, MNRAS, 325, 1603.
\bibitem[]{135}
Efstathiou, G. et al. (the 2dFGRS Team), 2002, MNRAS, 330, L29.
\bibitem[]{140}
Eisenstein, D.J. \& Hu, W., 1998, ApJ, 496, 605.
\bibitem[]{150}
Eisenstein, D.J. \& Zaldarriaga, M., 2001, ApJ, 546, 2.
\bibitem[]{153}
Eke, V.R., Cole, S. \& Frenk, C.S., 1996, MNRAS, 282, 263.
\bibitem[]{155}
Evrard, A.E. et al. (the Virgo Consortium), 2002, ApJ, 573, 7.
\bibitem[]{157}
Gazta\~naga, E., 1995, ApJ, 454, 561.
\bibitem[]{160}
Gazta\~naga, E. \& Baugh, C.M., 1998, MNRAS, 294, 229.
\bibitem[]{170}
Gnedin, N.Y. \& Hamilton, A.J.S., 2002, MNRAS, 334, 107.
\bibitem[]{180}
Hamilton, A.J.S. \& Tegmark, M., 2002, MNRAS, 330, 506.
\bibitem[]{195}
Hawkins, E. et al. (the 2dFGRS Team), 2002, astro-ph/0212375.
\bibitem[]{200}
Hoyle,F., Baugh, C.M., Shanks, T. \& Ratcliffe, A., 1999, MNRAS, 309, 659.
\bibitem[]{205}
Jaffe, A. H., et al. 2001, PhRvL, 86, 3475.
\bibitem[]{210}
Jarrett, T.H., Chester, T., Cutri, R., Schneider, S., Skrutskie, M.,
Huchra, J.P., 2000, AJ, 119, 2498.
\bibitem[]{215}
Jenkins, A. et al. (the Virgo Consortium), 1998, ApJ, 499, 20.
\bibitem[]{220}
Limber, D.N., 1954, ApJ, 119, 655.
\bibitem[]{230}
Lucy, L.B., 1974, AJ, 79, 745.
\bibitem[]{240}
Lucy, L.B., 1994, A\&A, 289, 983.
\bibitem[]{245}
Maddox, S.J., Efstathiou, G., Sutherland, W.J., Loveday,J., 1990, MNRAS, 
243, 692.
\bibitem[]{250}
Maddox, S.J., Efstathiou, G., Sutherland, W.J.,  1996, MNRAS, 283, 1227.
\bibitem[]{255}
Maller, A.H., McIntosh, D.H., Katz, N., \& Weinberg, M.D., 2003, 
submitted to ApJ, Astro-ph/0304005.
\bibitem[]{260}
McCracken, H.J., Le F\`ebre, O., Brodwin, M., Foucaud, S., Lilly, S.J.,
Crampton, D., Mellier, Y., 2001, A\&A, 376, 756. 
\bibitem[]{270}
Meiksin, A. \& White, M., 1999, MNRAS, 308, 1179.
\bibitem[]{280}
Netterfield, C.B., et al., 2002, ApJ, 571, 604.
\bibitem[]{290}
Norberg, P. et al. (the 2dFGRS Team), 2001, MNRAS, 328,64. 
\bibitem[]{300}
Norberg, P. et al. (the 2dFGRS Team), 2002b, MNRAS, 332, 827.
\bibitem[]{305}
Norberg, P. et al. (the 2dFGRS Team), 2002a, MNRAS, 336, 907.
\bibitem[]{307}
Ostriker, J.P., Steinhardt, P.J., 1995, Nature, 377, 600.
\bibitem[]{310}
Padilla, N.D., \& Baugh, C.M., 2002, MNRAS, 329, 431.
\bibitem[]{320}
Peacock, J.A., 1991, MNRAS, 253, 1.
\bibitem[]{325}
Peacock, J. A., 1997, MNRAS, 284, 885.
\bibitem[]{330}
Peacock, J.A. \& Dodds, S.J., 1996, MNRAS, 280, 19.
\bibitem[]{340}
Peebles, P.J.E., 1980, The Large-Scale Structure of the Universe.
Princeton Univ. Press, Princeton.
\bibitem[]{345}
Peebles, P.J.E., 1993, Principles of Physical Cosmology.
Princeton Univ. Press, Princeton.
\bibitem[]{350}
Percival, W.J. et al. (the 2dFGRS Team), 2001, MNRAS, 327, 1297. 
\bibitem[]{360}
Percival, W.J. et al. (the 2dFGRS Team), 2002, MNRAS, 337, 1068.
\bibitem[]{365}
Perlmutter, S. et al., 1999, ApJ, 517, 565.
\bibitem[]{367}
Saunders, W., Rowan-Robinson, M., Lawrence, A., 1992, MNRAS, 258, 134.
\bibitem[]{370}
Saunders, W. et al., 2000, MNRAS, 317, 55.
\bibitem[]{375}
Scoccimarro, R., Sheth, R.K., 2002, MNRAS, 329, 629.
\bibitem[]{377}
Seljak, U., Zaldarriaga, M., 1996, ApJ, 469, 437.
\bibitem[]{380}
Smith, R.E. et al. (the Virgo Consortium), 2002, astro-ph/0207664.
\bibitem[]{390}
Sutherland, W. et al., 1999, MNRAS, 308, 289.
\bibitem[]{400}
Szalay, A. et al. 2001, astro-ph/0107419.
\bibitem[]{410}
Tadros, H. \& Efstathiou, G., 1996, MNRAS, 282, 1381.
\bibitem[]{420}
Zaldarriaga, M., Scoccimarro, R. \& Hui, L., 2001, astro-ph/0111230.
\bibitem[]{425}
Zehavi et al., 2003, submitted to ApJ, astro-ph/0301280.
\bibitem[]{430}
Zucca, E. et al., 1997, A\&A, 326, 477.

\end{thebibliography}
\end{document}